\documentclass[amsmath,amssymb,showpacs,twocolumn,superscriptaddress,pr]{revtex4-1}
\usepackage{graphicx,bm,color,subfigure}

\begin{document}

\title{Chiral Spin Density Wave and $d+id$ Superconductivity in the Magic-Angle-Twisted Bilayer Graphene}

\author{Cheng-Cheng Liu}
\affiliation{School of Physics, Beijing Institute of Technology, Beijing 100081, China}

\author{Li-Da Zhang}
\affiliation{School of Physics, Beijing Institute of Technology, Beijing 100081, China}

\author{Wei-Qiang Chen}
\affiliation{Shenzhen Institute for Quantum Science and Engineering and Department of Physics, Southern University of Science and Technology, Shenzhen 518055, China}

\author{Fan Yang}
\email{yangfan\_blg@bit.edu.cn}
\affiliation{School of Physics, Beijing Institute of Technology, Beijing 100081, China}

\begin{abstract}
We model the newly synthesized magic-angle twisted bilayer-graphene superconductor with  two $p_{x,y}$-like Wannier orbitals on the superstructure honeycomb lattice, where the hopping integrals are constructed via the Slater-Koster formulism by symmetry analysis. The characteristics exhibited in this simple model are well consistent with both the rigorous calculations and experiment observations. A van Hove singularity and Fermi-surface (FS) nesting are found in the doping levels relevant to the correlated insulator and unconventional superconductivity revealed experimentally, base on which we identify the two phases as weak-coupling FS instabilities. Then, with repulsive Hubbard interactions turning on, we performed random-phase-approximation (RPA) based calculations to identify the electron instabilities. As a result, we find chiral $d+id$ topological superconductivity bordering the correlated insulating state near half-filling, identified as noncoplanar chiral spin-density wave (SDW) ordered state, featuring quantum anomalous Hall effect. The phase-diagram obtained in our approach is qualitatively consistent with experiments.
\end{abstract}



\maketitle

\textit{\textcolor{blue}{Introduction.---}} The newly  revealed ``high-temperature superconductivity (SC)"\cite{SC} in the ``magic-angle" twisted bilayer-graphene (MA-TBG) has caught great research interests\cite{Volovik2018,Roy2018,Po2018,Xu2018,Yuan2018,Baskaran2018,Phillips2018,Kivelson2018}. In such a system, the low energy electronic structure can be dramatically changed by the twist. It was shown that some low energy flat bands, which are well separated with other high energy bands, appear when the twist angle is around $1.1^{\circ}$. A correlated insulating state is observed when the flat bands are near half-filled \cite{Mott}. Doping this correlated insulator leads to SC with highest critical temperature $T_c$ up to 1.7 K. This system looks similar to the cuprates in terms of phase diagram and the high ratio of $T_c$ over the Fermi-temperature $T_F$. In fact, it was argued that the insulating state was a Mott insulator, while the MA-TBG was an analogy of cuprate superconductor. Since the structure of the MA-TBG is  in situ tunable, it was proposed that this system can serve as a good platform to study the pairing mechanism of the high-$T_c$ SC, the biggest challenge of condensed-matter physics.

However, the viewpoint that the SC in MA-TBG is induced by doping a Mott-insulator suffers from the following three inconsistencies with experimental results. Firstly, the so-called ``Mott-gap" extrapolated from the temperature-dependent conductance is just about 0.31 meV\cite{Mott}, which is much lower than the band width of the low energy emergent flat bands ($\sim$10 meV). Such a tiny ``Mott-gap" can hardly be consistent with the ``Mott-physics". Secondly, the behaviors upon doping into this insulating phase is different from those for  doping a Mott-insulator, as analyzed in the following for the positive filling as an example. In the case of electron doping with respect to the half-filling, the system has small Fermi pocket with area proportional to doping, which is consistent with a doped Mott insulator\cite{SC}. However, in the hole doping case, slightly doping leads to a large Fermi surface (FS) with area proportional to the electron concentration of the whole bands instead of the hole concentration with respect to the half-filling\cite{SC, Mott}. Such behavior obviously conflicts with the ``Mott-physics". Thirdly, some samples which exhibit the so-called ``Mott-insulating" behavior at high temperature become SC upon lowering the temperature\cite{Notice}. Such a behavior is more like to be caused by the competing between SC and some other kind of orders, such as density waves, instead of ``Mott physics".

In this Letter, we study the problem from weak coupling approach, wherein electrons on the FS acquire effective attractions through exchanging spin fluctuations, which leads to Cooper pairing. After analyzing the characteristics of the low energy emergent band structure, an effective $p_{x,y}$-orbital tight-binding model\cite{Yuan2018} on the emergent honeycomb lattice is adopted, but with the hopping integrals newly constructed via the Slater-Koster formulism\cite{Slater1954}, which is re-derived based on the symmetry of the system (Supplementary Material I\cite{SupplMater}). The characteristics of the constructed band structure is qualitatively consistent with both the rigorous multi-band tight-binding results\cite{Nguyen2017,Moon2012} and experiments\cite{SC,Mott}. Moreover the band degeneracy at high-symmetric points or lines is compatible with the corresponding irreducible representations\cite{Yuan2018}. Then after the Hubbard-Hund interaction turns on, we performed RPA based calculations to study the electron instabilities. Our results identify the correlated insulator near half-filling as FS-nesting induced noncoplanar chiral SDW insulator, featuring quantum anomalous Hall effect (QAHE). Bordering this SDW insulator is chiral $d+id$ topological superconducting state. The obtained phase diagram is qualitatively consistent with experiments.

\textit{\textcolor{blue}{The model.---}} For the MA-TBG, the small twist angle between the two graphene layers causes Moire pattern which results in much enlarged unit cell, consequently thousands of energy bands are taken into account\cite{Nguyen2017,Moon2012},  and the low-energy physics are dramatically changed\cite{Nguyen2017,Moon2012,Fang2015,Santos2007,Santos2012,Shallcross2008,Bistritzer2011,Bistritzer2010,Uchida2014,Mele2011,Mele2010,Sboychakov2015,Morell2010,Trambly2010,Latil2007,Trambly2012,Gonza2013,Luis2017,Cao2016,Ohta2012,Kim2017,Huder2018,Li2017}.
Remarkably, four low energy nearly-flat bands with a total bandwidth of about 10 meV emerge which are well isolated from the high energy bands. Since both the correlated insulating and the superconducting phases emerge when these low energy bands are partially filled, it's urgent to provide an effective model with relevant degrees of freedom to capture the low energy band structure.

\begin{figure}[htbp]
\centering
\includegraphics[width=0.48\textwidth]{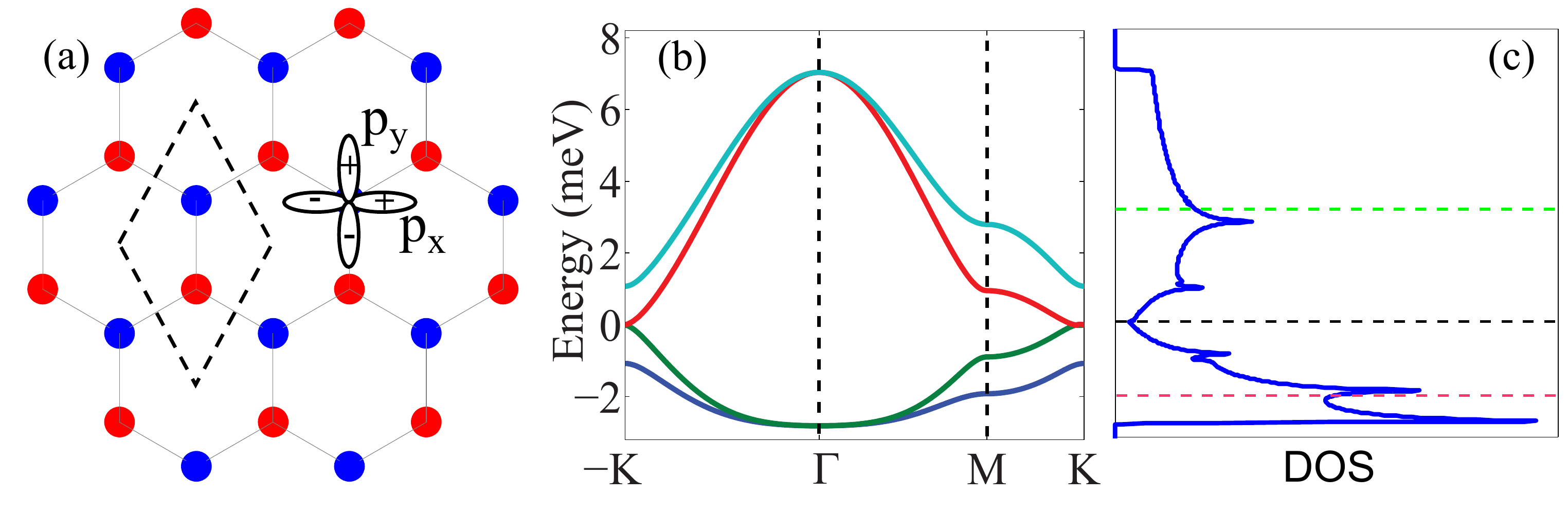}
\caption{(a) Schematic diagram for our model. The dashed rhombus labels the unit cell of the emergent honeycomb lattice with the $p_{x},p_{y}$-like Wannier orbitals on each site. (b) Band structure and (c) DOS of MA-TBG. The red, black and green horizontal dashed lines in (c) label the filling $\delta$ of $-\frac{1}{2}$, 0 and $\frac{1}{2}$ respectively. The Slater-Koster parameters $t_{\sigma/\pi}^{ij}$ are chosen as $t^{(1)}_{\sigma}=2$ meV, $t^{(1)}_{\pi}=t^{(1)}_{\sigma}/1.56$, $t^{(2)}_{\sigma}=t^{(1)}_{\sigma}/7$, and $t^{(2)}_{\pi}=t^{(2)}_{\sigma}/1.56$. Here the superscript (1)/(2) represents NN or NNN bondings respectively.}\label{band}
\end{figure}

By analyzing the degeneracy and representation of the flat bands at all three of the high symmetry points $\Gamma$, $K$ and $M$, a honeycomb lattice rather than the triangular one should be adopted to model the low-energy physics of MA-TBG\cite{Po2018,Yuan2018}.
The emergent honeycomb lattice consists of two sublattices originating from different layers. Further symmetry analysis shows the related Wannier orbitals on each site belong to the $p_x$ and $p_y$ symmetry\cite{Yuan2018}.
Therefore, we can construct the hopping integrals between the $p_{x,y}$-like orbitals on the honeycomb lattice via the Slater-Koster formulism\cite{Slater1954} based on symmetry analysis\cite{SupplMater}, which reflects coexisting $\sigma$ and $\pi$ bondings\cite{Wu2008a,Wu2008b,Zhang2014,Liu2014,Yang2015}. Our tight-binding (TB) model up to the next nearest neighbor (NNN) hoping thus obtained reads,
\begin{equation}\label{tb}
H_{tb}=\sum_{i\mu,j\nu,\sigma}t_{i\mu,j\nu}c_{i\mu\sigma}^{\dagger}c_{j\nu\sigma}-\mu_c\sum_{i\mu\sigma}c_{i\mu\sigma}^{\dagger}c_{i\mu\sigma}.
\end{equation}
Here $\mu,\nu=x,y$ represent the $p_{x},p_{y}$ orbitals shown in Fig.~\ref{band}(a),  $i,j$ stand for the site and $\mu_c$ is the chemical potential determined by the filling $\delta\equiv n/n_s-1$ relative to charge neutrality. Here $n$ is the average electron number per unit cell, $n_s=4$ is the $n$ for charge neutrality. The hopping integral $t_{i\mu,j\nu}$ can be obtained as
\begin{equation}\label{slater_koster}
t_{i\mu,j\nu}=t_{\sigma}^{ij}\cos\theta_{\mu,ij}\cos\theta_{\nu,ij}+t_{\pi}^{ij}\sin\theta_{\mu,ij}\sin\theta_{\nu,ij},
\end{equation}
where $\theta_{\mu,ij}$ denotes the angle from the direction of $\mu$ to that of $\mathbf{r}_{j}-\mathbf{r}_{i}$.
The Slater-Koster parameters $t_{\sigma/\pi}^{ij}$ represent the hopping integrals contributed by $\sigma/\pi$- bondings. More details about the band structure are introduced in Supplementary Materials II\cite{SupplMater}.

The band structure of our TB model (\ref{tb}) is shown in Fig.~\ref{band}(b). Investigating the degeneracy pattern here, one finds that the $\Gamma$-point is doubly degenerate for each two bands, the $M$-point is non-degenerate, and as for the two $K$-points, both degenerate Dirac crossing and non-degenerate splitting (Dirac mass generation) coexist. Such degeneracy pattern is consistent with both symmetry representation\cite{Yuan2018,Po2018} and rigorous results\cite{Nguyen2017,Moon2012}. Note that the Dirac mass generation is important\cite{Yuan2018} in understanding the quantum oscillation experiment, wherein 4-fold Landau level degeneracy (including spin-degeneracy) is observed near the charge neutrality\cite{SC}. The model parameters $t_{\sigma/\pi}^{ij}$ (introduced in the caption of Fig.~\ref{band}(a)) are tuned so that the renormalized Fermi velocity (which is about $\frac{1}{25}$ of that of monolayer graphene), as well as the total band width (about 10meV), are consistent with experiments. Note that due to the presence of NNN-hoppings, the band structure is particle-hole asymmetric with the negative (n-) energy part flatter than the positive (p-) one, consistent with experiment\cite{SC}. The density of state (DOS) shown in Fig.~\ref{band}(c) suggests that the $\pm\frac{1}{2}$ fillings relevant to experiments are near the Van-Hove (VH) fillings $\delta_V$ ($\approx\pm 0.425$) introduced below.
\begin{figure}[htbp]
\centering
\includegraphics[width=0.48\textwidth]{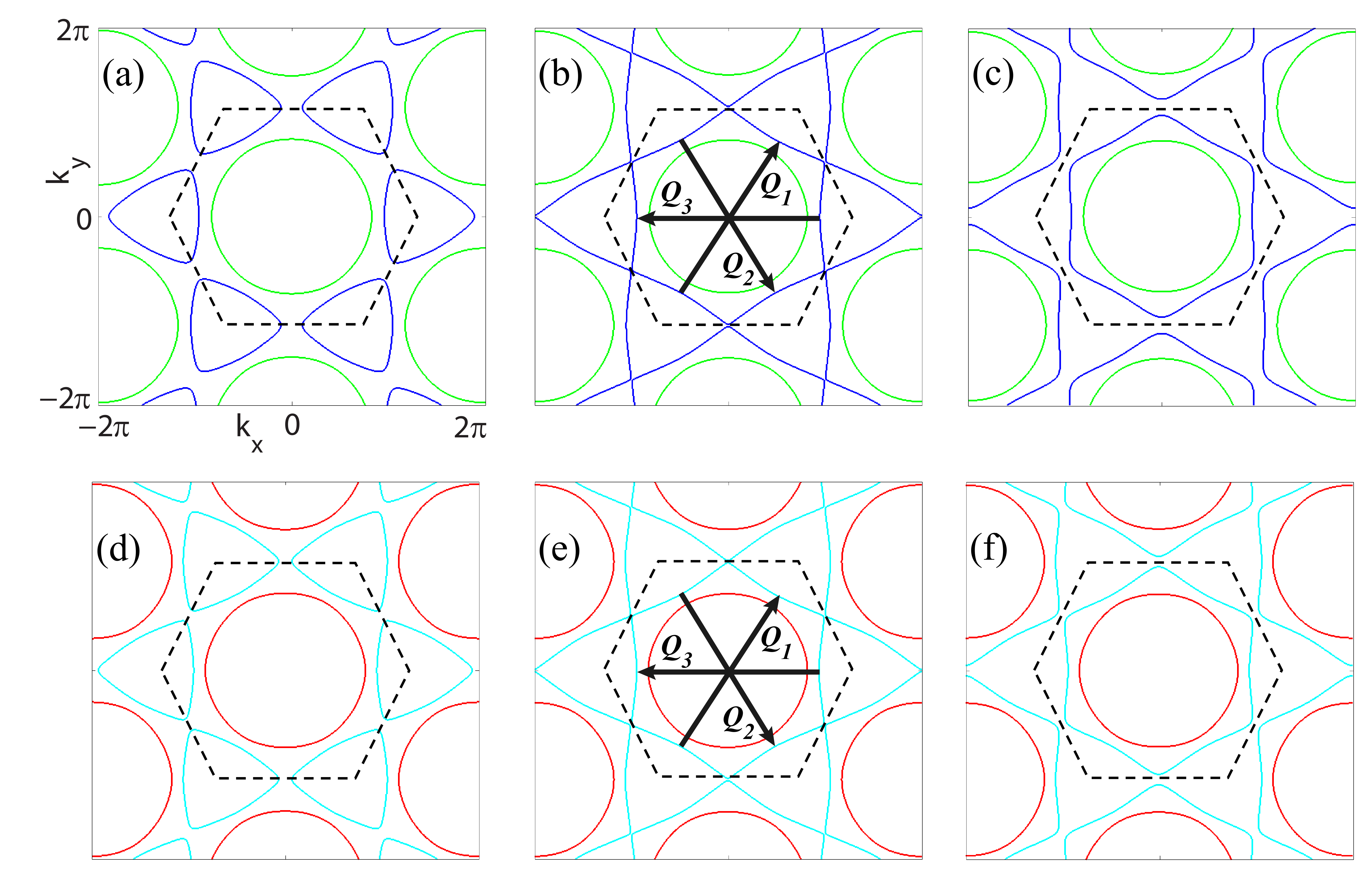}
\caption{The evolution of FS. (a)-(c) FS at fillings -0.40, -0.425 ($\delta_V$) and -0.45. (d)-(f) FS at fillings 0.41, 0.425 ($\delta_V$) and 0.44. The central hexagons in every plot in black dashed line indicate the Brillouin Zone. The FS-nesting vectors $\bm{Q}_{\alpha}$ $(\alpha=1,2,3)$ are marked in (b) and (e).}\label{FS}
\end{figure}

The evolution of the FS with filling near $\pm\frac{1}{2}$ is shown in Fig.~\ref{FS}. One finds that in both the n- and p- parts, a Lifshitz transition takes place at some critical fillings $\delta_V$, which changes the FS topology. At the Lifshitz transition point, there is VH singularity (VHS) at the three $M$-points, and the FS is nearly-perfect nested, as shown in Fig.~\ref{FS}(b) and (e), with the three marked nesting vectors $\bm{Q}_{\alpha}$ $(\alpha=1,2,3)$ also near the three $M$-points. One finds that the situation of FS-nesting is asymmetric before and after the Lifshitz transition: before $\delta_V$, there are three FS patches with bad nesting; after $\delta_V$, two FS patches are left with the outer one well nested. This asymmetry on FS-nesting is closely related to the asymmetry in the phase diagram studied below. Note that the $|\delta_V|$ for the p- and n- parts are only approximately equal. As shown in Supplementary Materials II\cite{SupplMater}, these characteristics don't obviously change with model parameters in reasonable range and $\delta_V$ is generally near $\pm\frac{1}{2}$.

It's proposed here that the SC detected in the MA-TBG is driven not by electron-phonon coupling but by electron-electron interactions (See Supplementary Materials III\cite{SupplMater} for the analysis). We adopt the following repulsive Hubbard-Hund model proposed in Ref\cite{Yuan2018},
\begin{eqnarray}\label{model}
H=&H_{tb}&+H_{int}\nonumber\\
H_{int}=&U&\sum_{i\mu}n_{i\mu\uparrow}n_{i\mu\downarrow}+V\sum_{i}
n_{ix}n_{iy}+J_{H}\sum_{i}\Big[\sum_{\sigma\sigma^{\prime}}\nonumber\\&c^{\dagger}_{ix\sigma}&c^{\dagger}_{iy\sigma^{\prime}}
c_{ix\sigma^{\prime}}c_{iy\sigma}+(c^{\dagger}_{ix\uparrow}c^{\dagger}_{ix\downarrow}
c_{iy\downarrow}c_{iy\uparrow}+h.c.)\Big]
\end{eqnarray}
where $U=V+2J_H$. Adopting $U=1.5$ meV, we have considered both $J_H=0.1U$ and $J_H=0$, with the two cases giving qualitatively the same results. Strictly speaking, the Hubbard-Hund interactions $H_{int}$ describing the atomic interactions do not apply here for our extended effective orbitals. However, as will be seen, the electron instabilities here are mainly determined by the VHS and the FS-nesting, and will not be strongly affected by the concrete formulism of the interactions, as long as it's repulsive. Therefore, the model(\ref{model}) can be a good start point.

\textit{\textcolor{blue}{Electron instabilities and phase diagram.---}} We adopt the standard multi-orbital RPA approach\cite{RPA1,RPA2,RPA3,RPA4} to study the electron instabilities of the system. We start from the normal-state susceptibilities in the particle-hole channel and consider its renormalization due to interactions up to the RPA level. Through exchanging spin or charge fluctuations represented by these susceptibilities, the electrons near the FS acquire effective attractions, which leads to pairing instability for arbitrarily weak interactions. However, when the repulsive interaction strength $U$ rises to some critical value $U_c$, the spin susceptibility diverges, which leads to SDW order. More details can be found in Supplementary Materials III\cite{SupplMater}.

\begin{figure}[htbp]
\centering
\includegraphics[width=0.48\textwidth]{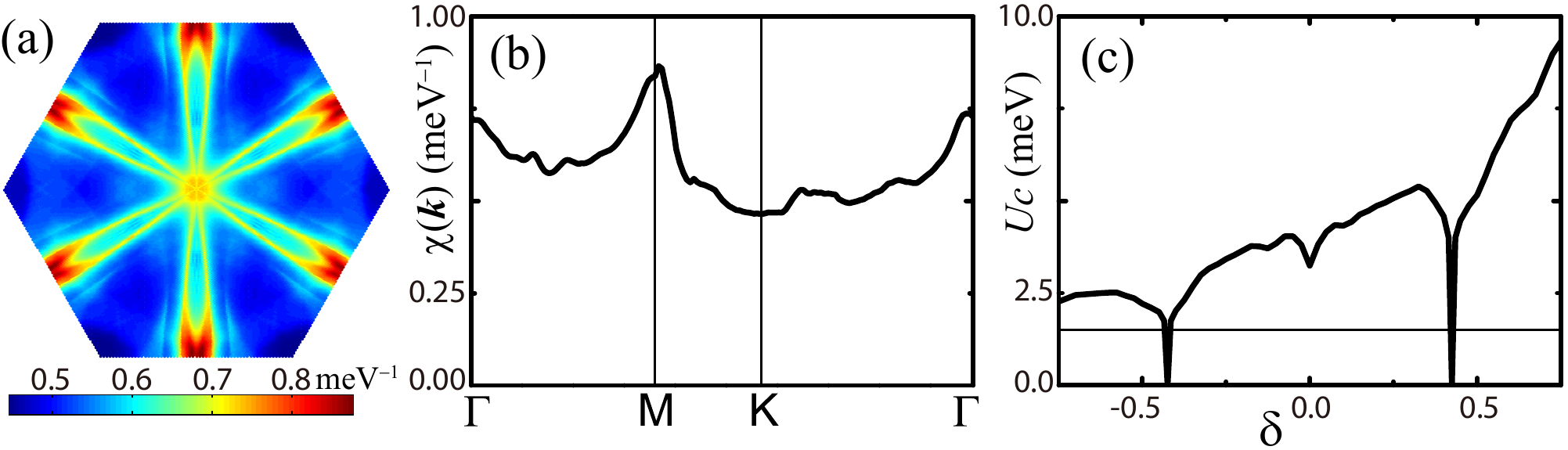}
\caption{Distribution of $\chi(\bm{k})$ for $\delta\to\delta_V=-0.425$ (a) in the Brillouin Zone, and (b) along the high-symmetric lines. (c) The filling dependence of $U_c$ (for $J_H=0.1U$), with the horizontal line representing $U=1.5$ meV adopted in our calculations.}\label{magnet}
\end{figure}

The bare susceptibility tensor is defined as,
\begin{align}\label{chi0}
\chi^{(0)l_1l_2}_{l_3l_4}(\bm{k},\tau)\equiv
&\frac{1}{N}\sum_{\bm{k}_1\bm{k}_2}\left\langle
T_{\tau}c_{l_1}^{\dagger}(\bm{k}_1,\tau)
c_{l_2}(\bm{k}_1+\bm{k},\tau)\right.                      \nonumber\\
&\left.\times c_{l_3}^{\dagger}(\bm{k}_2+\bm{k},0)
c_{l_4}(\bm{k}_2,0)\right\rangle_0.
\end{align}
Here $\langle\cdots\rangle_0$ denotes the thermal average for the non-interacting system, and $l_{i(i=1,...,4)}=1,2,3,4$ are the sublattice-orbital indices. Fourier transformed to the imaginary frequency space, the obtained $\chi^{(0)l_1,l_2}_{l_3,l_4}(\bm{k},i\omega_n)$ can be taken as a matrix with $l_1l_2/l_3l_4$ to be the row/column indices. The largest eigenvalue $\chi(\bm{k})$ of the zero-frequency susceptibility matrix $\chi^{(0)l_1l_2}_{l_3l_4}(\bm{k},i\omega_n=0)$ as function of $\bm{k}$ for $\delta\to\delta_V=-0.425$ is shown in the whole Brillouin Zone in Fig.~\ref{magnet}(a), and along the high-symmetry lines in Fig.~\ref{magnet}(b).

Figure~\ref{magnet}(a) and (b) show that the distribution of $\chi(\bm{k})$ for $\delta\to\delta_{V}$ peaks near the three $M$- points in the Brillouin Zone, which originates from the FS-nesting shown in Fig.~\ref{FS}(b). In the thermal-dynamic limit, these peaks will diverge due to diverging DOS. Therefore arbitrarily weak interactions will induce density-wave type of instability. The RPA treatment shows that repulsive interactions suppress the charge susceptibility $\chi^{(c)}$ and enhance the spin susceptibility $\chi^{(s)}$(Supplementary Materials III\cite{SupplMater}). Therefore, SDW order emerges for arbitrarily weak Hubbard interactions in our model at $\delta_{V}$. We identify the correlated insulator observed by experiment\cite{Mott} to be the SDW insulator proposed here at $\delta_V$, which is near $\pm\frac{1}{2}$.

Note that the competition among the three-fold degenerate FS-nesting vectors $\bm{Q}_{\alpha}$ $(\alpha=1,2,3)$ will drive noncoplanar SDW order with spin chirality, featuring QAHE\cite{TaoLi,Martin,Kato,Ying}. To clarify this point, let's extrapolate the eigenvectors $\xi(\bm{Q_\alpha})$ corresponding to the largest eigenvalue of the susceptibility matrix $\chi^{(0)}(\bm{k},i\omega_n=0)$ for $\bm{k}\to \bm{Q_\alpha}$. Defining the magnetic order parameters $\bm{S}_{i\mu\nu}\equiv \left\langle c^{\dagger}_{i\mu s}\bm{\sigma}_{s s^{\prime}}c_{i\nu s^{\prime}}\right\rangle$, the divergence of $\chi(\bm{Q}_{\alpha})$ requires spontaneous generation of magnetic order with $\bm{S}_{i\mu\nu}\propto\xi_{\mu\nu}(\bm{Q_\alpha})e^{i\bm{Q}_\alpha\cdot\bm{R}_i}\bm{n}_\alpha$, with the global unit vector $\bm{n}_\alpha$ pointing anywhere. Now we have three degenerate $\bm{Q}_\alpha$, which perfectly fit into the three spatial dimensions: $\bm{S}_{i\mu\nu}\propto(\xi_{\mu\nu}(\bm{Q_1})e^{i\bm{Q}_1\cdot\bm{R}_i},\xi_{\mu\nu}(\bm{Q_2})e^{i\bm{Q}_2\cdot\bm{R}_i},
\xi_{\mu\nu}(\bm{Q_3})e^{i\bm{Q}_3\cdot\bm{R}_i})$. Such noncoplanar SDW order with spin chirality may lead to nontrivial topological Chern-number in the band structure, resulting in QAHE\cite{TaoLi,Martin,Kato,Ying}.

When the filling is away from $\delta_V$, SDW order only turns on when $U>U_c$, where the renormalized spin susceptibility tensor $\chi^{(s)}$ diverges. The filling-dependence of $U_c$ for $J_H=0.1U$ is shown in Fig.~\ref{magnet}(c) (the case of $J_H=0$ yields similar result), where SDW order is only present within a narrow regime centering at the two $\delta_V$. When $U<U_c$, through exchanging short-ranged spin and charge fluctuations between a Cooper pair, an effective pairing interaction vertex $V^{\alpha\beta}(\bm{k},\bm{k}')$  will be developed, which leads to the following linearized gap equation near $T_c$(See Supplementary Materials III\cite{SupplMater}):
\begin{align}\label{gapeq}
-\frac{1}{(2\pi)^2}\sum_{\beta}\oint_{FS}
dk'_{\Vert}\frac{V^{\alpha\beta}(\bm{k},\bm{k}')}
{v^{\beta}_{F}(\bm{k}')}\Delta_{\beta}(\bm{k}')=\lambda
\Delta_{\alpha}(\bm{k}).
\end{align}
Here $\beta$ labels the FS patch, $v^{\beta}_F(\bm{k}')$ is the Fermi velocity, and $k'_{\parallel}$ is the tangent component of $\bm{k}'$. Eq.(\ref{gapeq}) becomes an eigenvalue problem after discretization, with the eigenvector $\Delta_{\alpha}(\bm{k})$ representing the gap form factor and the eigenvalue $\lambda$ determining corresponding $T_c$ through $T_c\propto e^{-1/\lambda}$. From symmetry analysis (See Supplementary Materials IV\cite{SupplMater}), each solved $\Delta_{\alpha}(\bm{k})$ is attributed to one of the 5 irreducible representations of the $D_6$ point-group of our model (or $D_3$ point-group for real material with spin-SU(2) symmetry), which corresponds to $s,(p_{x},p_{y}),(d_{x^2-y^2},d_{xy}),f_{x^3-3xy^2},f^{\prime}_{y^3-3yx^2}$ wave pairings respectively. Note that only intra-band pairing is considered here.
\begin{figure}[htbp]
\centering
\includegraphics[width=0.48\textwidth]{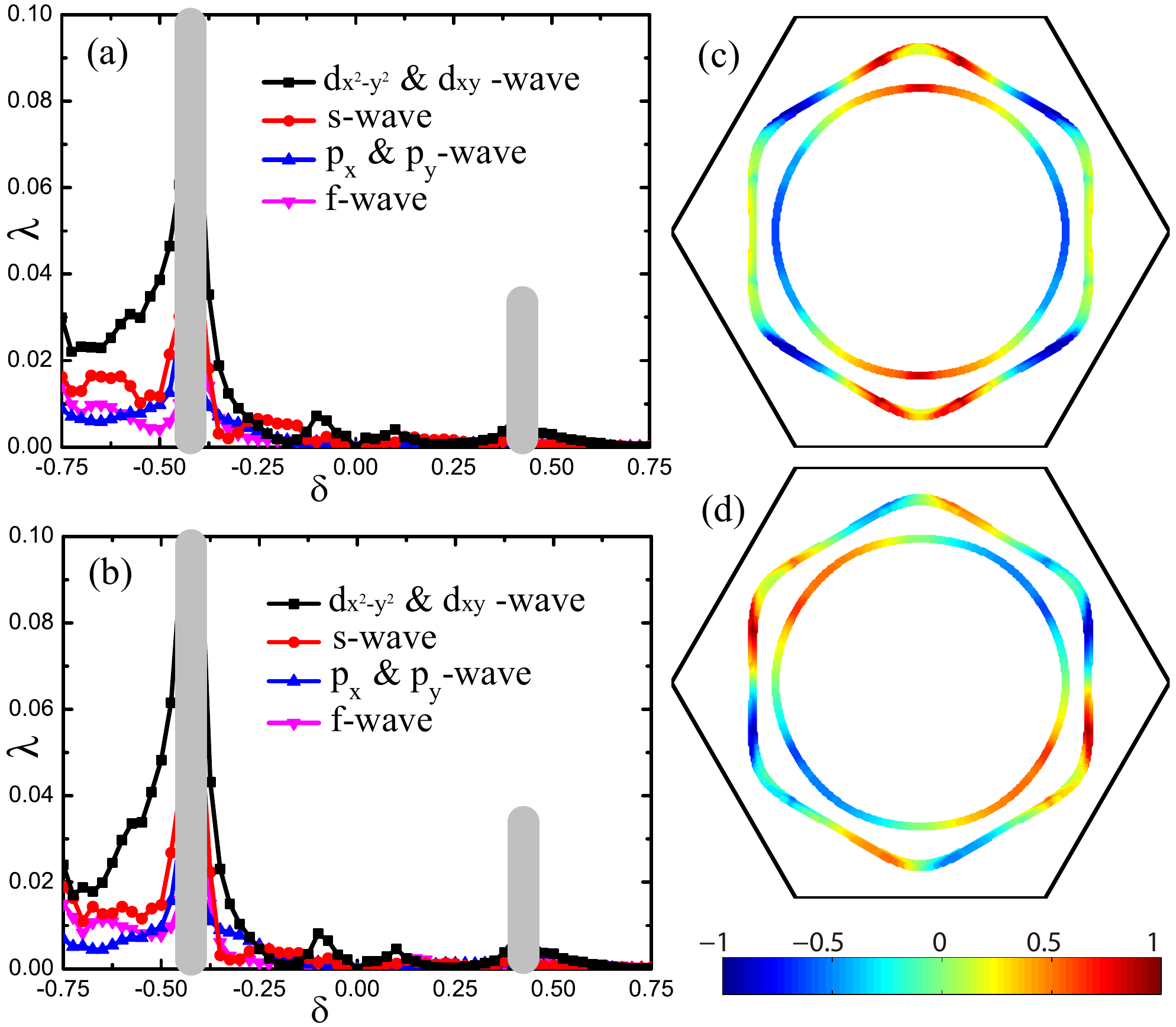}
\caption{The doping dependence of the largest eigenvalues $\lambda$ of all pairing symmetries for (a) $J_H=0.1U$ and (b) $J_H=0$. Note that the shown eigenvalue for the $f$-wave is the larger one of the two different f-symmetries. The vertical bold grey lines indicate the SDW regime. (c) and (d) are the gap functions of $d_{x^2-y^2}$ and $d_{xy}$-wave symmetries at doping $\delta=-0.5$, respectively.}\label{SC}
\end{figure}

The filling-dependence of the largest pairing eigenvalue for each pairing symmetry in the superconducting regimes is shown in Fig.~\ref{SC}(a) for $J_H=0.1U$ and (b) for $J_H=0$, in together with the SDW regimes.No obvious difference is found between  Fig.~\ref{SC}(a) and (b). Figure~\ref{SC}(a) or (b) can also be viewed as the phase diagram, which exhibits the following three remarkable features. Firstly, although SDW order induced by FS-nesting is present at $\delta_V$ for both p- and n- fillings, the SC order is only obvious on the n- part, because the bands in the n- part is flatter, which leads to higher DOS. Secondly, the SC order is strong near the SDW regime, as the spin-fluctuation there is strong. This feature makes the system look similar to the cuprates. Note that the unphysical divergence of $\lambda$ ($T_c$ will be very high) just bordering the SDW regime is only an artifact of the RPA treatment, which can be eliminated through introducing self-consistent correction to the single-particle Green's function, as done in the FLEX approach\cite{FLEX1,FLEX2,FLEX3}. Thirdly, the SC order on the left side of $\delta_V$ is stronger than that on its right side for the n- part. This asymmetry is due to the asymmetry in FS-nesting: the FS in Fig.~\ref{FS}(c) (the left side of $\delta_V$) is better nested than that in Fig.~\ref{FS}(a)(the right side of $\delta_V$). All these features are well consistent with experiments.

From Fig.~\ref{SC}(a) or (b), the leading pairing symmetry near $\delta_V$ relevant to experiment is the degenerate $(d_{x^2-y^2}, d_{xy})$ doublets, with their gap form factors shown in Fig.~\ref{SC}(c) and (d). While the gap function of $d_{x^2-y^2}$ depicted in Fig.4(c) is symmetric about the x- and y- axes in the reciprocal lattice, that of $d_{xy}$ depicted in Fig.4(d) is asymmetric about them. This singlet pairing symmetry is driven by the antiferromagnetic spin-fluctuations here. Physically, the key factors which determine the formation of d-wave SC are the VHS and FS-nesting. Firstly, the VHS takes place at the three time-reversal-invariant momenta, which only supports singlet pairings\cite{Hongyao}. Secondly, the location of the VHS and the FS-nesting vectors on the outer FS here at $\delta_V$ is nearly the same as those of the single-layer graphene at quarter doping\cite{Nandkishore2012}. Then from the renormalization group (RG) analysis\cite{Nandkishore2012,Wang2012,Nandkishore2014}, both systems should share the same pairing symmetry, i.e. the degenerate d-wave. Therefore, the d-wave SC here is mainly determined by the features of the FS, and depends little on the details of the repulsive interactions.

The degenerate $(d_{x^2-y^2}, d_{xy})$ doublets will further mix into fully-gapped $d_{x^2-y^2}\pm i d_{xy}$ ($d+id$) SC in the ground state to minimize the energy (See Supplementary Materials V\cite{SupplMater}). This chiral pairing state breaks time-reversal-symmetry and belongs to class C topological SC\cite{Schnyder}, characterized by integer topological quantum number Z and thus can host topologically protected boundary fermion modes, which appeals for experimental verification.

\textit{\textcolor{blue}{Discussion and Conclusion.---}} Note that there are two FS patches at $\delta_V$ in our model, with only the outer one well nested, as shown in Fig.~\ref{FS}(b) and (e). Thus for weak $U$, the FS-nesting driven SDW order can only gap out the outer FS, leaving the inner pocket untouched. However, as argued in Ref\cite{SC}, the interaction strength in the MA-TBG is not weak in comparison with the bandwidth. Therefore, the SDW order might be strong enough to touch and gap out the inner pocket as well, leading to a tiny net gap on that pocket, which might be related to the so-called ``Mott-gap" of 0.31 meV detected by experiment\cite{Mott}. Moreover, this gap caused by noncoplanar SDW order will easily be closed by Zeeman-coupling to an applied external magnetic field\cite{Kato}, driving the metal-insulator transition detected by experiment\cite{SC}.

The asymmetry on the situation of FS-nesting on different doping sides of $\delta_V$ shown in Fig.2 might also be related to the asymmetry observed in quantum oscillation experiments\cite{SC}. As the FS for the side $|\delta|\gtrsim|\delta_V|$ is better nested than that for the other side, it's possible for some range of $U$ that SDW order only emerges on that side, in which case a small pocket with the area proportional to $|\delta|-|\delta_V|$ only emerges on the side $|\delta|\gtrsim|\delta_V|$, consistent with quantum oscillation experiments\cite{SC}.

We notice a peak in the DOS near the band bottom, as shown in Fig.~\ref{band}(c). Careful investigation into the band structure reveals that the peak is caused by band flattening near the bottom. The FS there only includes two small pockets, and no VHS or FS-nesting can be found, as shown in Supplementary Materials II\cite{SupplMater}. In this region, ferromagnetic metal caused by Stoner criteria instead of SC might be developed.

In conclusion, our adopted effective $p_{x,y}$-orbital TB model on the emergent honeycomb lattice with the newly constructed hopping integrals well captures the main characteristics of real system. Remarkably, Lifshitz transitions take place at VH fillings near $\pm\frac{1}{2}$. The VHS and the FS-nesting with three-fold degenerate nesting vectors drive the system into noncoplanar chiral SDW order, featuring QAHE. Bordering the chiral SDW phase is $d+id$ chiral topological SC (TSC) driven by short-ranged antiferromagnetic fluctuations. The phase diagram of our model is similar with experiments. The MA-TBG thus might provide the first realization of the intriguing $d+id$ chiral TSC proposed previously\cite{Baskaran,Black,Gonzalez,Nandkishore2012,Wang2012,Black2012,Liu2013,Nandkishore2014} in graphene-related systems.

\section*{Acknowledgements}
 We are grateful to the helpful discussions with Fu-Chun Zhang, Yuan-Ming Lu, Qiang-Hua Wang, Yi-Zhuang You, Noah Fan-Qi Yuan, Xian-Xin Wu, Hong Yao, Zheng-Cheng Gu, Yue-Hua Su, Yi Zhou, Tao Li and Yugui Yao. This work is supported by the NSFC (Grant Nos. 11674025, 11774028, 11604013, 11674151, 11334012, 11274041), the National Key Research and Development Program of China (No. 2016YFA0300300), Beijing Natural Science Foundation (Grant No. 1174019), and Basic Research Funds of Beijing Institute of Technology (No. 2017CX01018). {\color{blue}{\underline {Cheng-Cheng}} {\underline {Liu and Li-Da Zhang contributed equally to this work.}}}

\newpage

\begin{widetext}

\section{Appendix:Re-derivation of the tight-binding model with explicit parameters
by symmetry}

The twisted bilayer-graphene has the $D_{3}$ point-group symmetry.
The emergent honeycomb superstructure for MA-TBG is slightly buckled, as shown in Fig.\ref{Fig_TB_symmetry}. The lattice constant of the emergent honeycomb lattice is about 12.8 nm, much larger than its buckling height (0.335 nm). As a result, the emergent honeycomb lattice approximately has $D_{6}$ point-group symmetry.

The two relevant orbitals consist of two Wannier orbitals with x/y symmetry, recorded as $\phi_{x},\phi_{y}$,
which can be taken as the basis functions of the two-dimensional (2D) irreducible representations
$E$ of point-group $D_{3}$.

\begin{figure}
\includegraphics[scale=0.7]{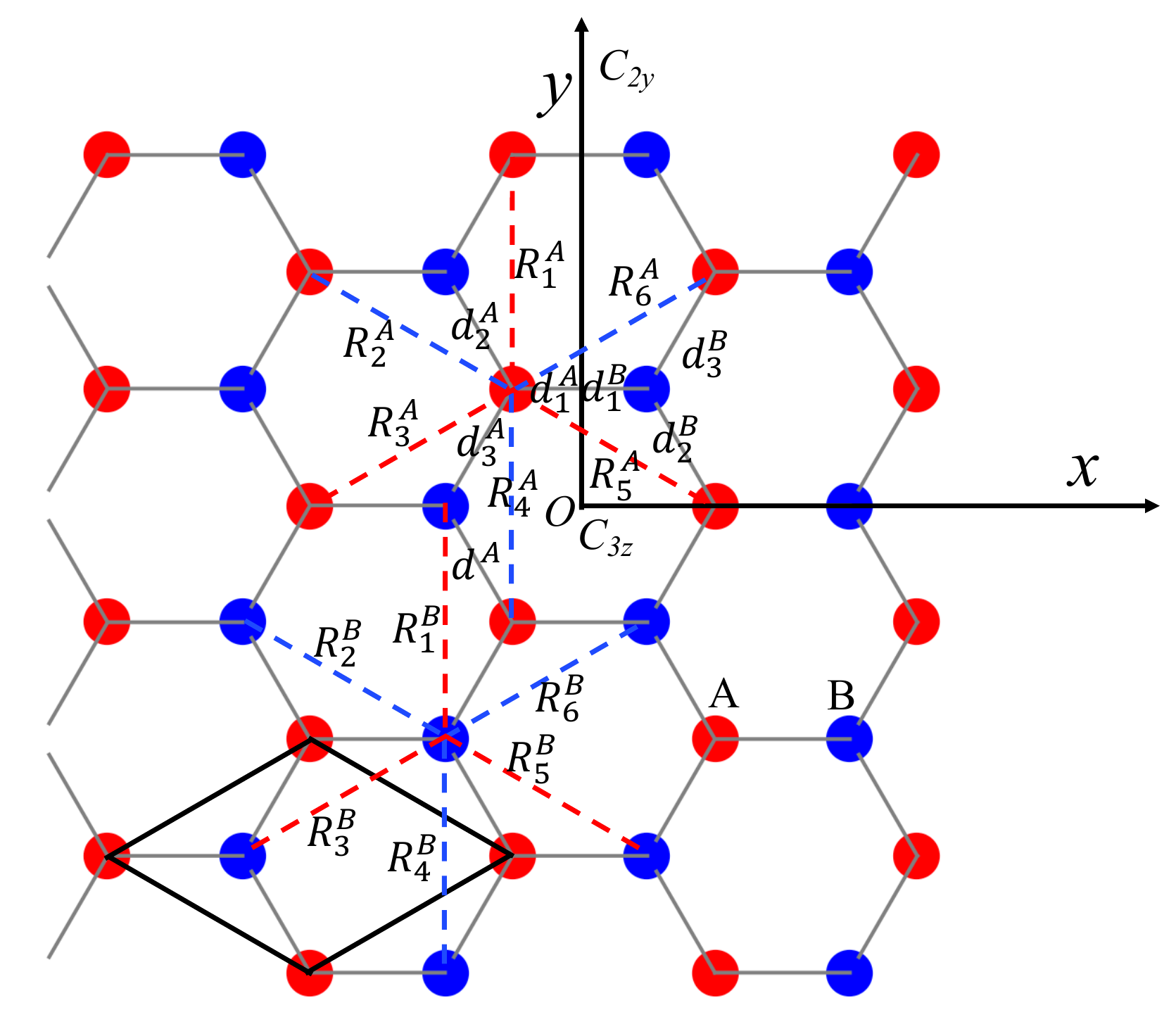}
\caption{The emergent honeycomb superstructure for MA-TBG and the various neighbors. A, B label the two sublattices, $d_{i=1-3}^{A,B}$ denote the three nearest neighboring vectors from A or B,  and $R_{i=1-6}^{A,B}$ denote the six next nearest neighboring vectors from A or B.
The black rhombus shows the two-dimensional (2D) unit cell for the emergent honeycomb superstructure.}\label{Fig_TB_symmetry}
\end{figure}

The matrix elements of the Hamiltonian in momentum space can be
easily obtained as
\begin{equation}
H_{\mu\nu}\left(k\right)=\sum_{\mathbf{\delta}}{\displaystyle e^{i\mathbf{k}\cdot\mathbf{\delta}}E_{\mu\nu}\left(\mathbf{\delta}\right)},
\end{equation}
with the hopping integral between the orbitals $|\phi_{\mu}>$ at site $\mathbf{0}$ and $|\phi_{\nu}>$ at site $\mathbf{\delta}$ defined
as
\begin{equation}
E_{\mu\nu}\left(\mathbf{\delta}\right)\equiv<\phi_{\mu}\left(\mathbf{r}\right)|\hat{H}|\phi_{\nu}\left(\mathbf{r}-\delta\right)>.
\end{equation}
Given $E_{\mu\nu}\left(\mathbf{\delta}\right)$, the hopping integrals between site $\mathbf{0}$ and all the other symmetry-related sites $\hat{g}\mathbf{\delta}$ can be obtained by
\begin{equation}
E_{\mu\nu}\left(\hat{g}\mathbf{\delta}\right)=D\left(\hat{g}\right)E_{\mu\nu}\left(\mathbf{\delta}\right)\left[D\left(\hat{g}\right)\right]^{\dagger},
\end{equation}
where $\hat{g}$ is the symmetry operator of point-group $D_{3}$,
$D\left(\hat{g}\right)$ is the $2\times2$ representation matrix
of the 2d irreducible representations $E$. Considering the sublattice
degree of freedom, the final Hamiltonian is $4\times4$ matrix. As
shown in Fig.\ref{Fig_TB_symmetry}, the point-group $D_{3}$ has the generators: three-fold
rotation along $\mathbf{z}$ axis $C_{3z}$ and two-fold rotation
along $\mathbf{y}$ axis $C_{2y}$.

Firstly, let's consider the nearest-neighbor (NN) hopping integrals. The $A$ site has three NN $B$ sites, with the corresponding vectors $\vec{AB}$ labeled
by $\mathbf{d}_{i=1-3}^{A}$. By the symmetry operator $C_{3z}$,
$\mathbf{d}_{1}^{A}$ is rotated to $\mathbf{d}^{A}$, which is equivalent
to $\mathbf{d}_{2}^{A}$. Similarly, $\mathbf{d}_{1}^{A}$ is rotated
to $\mathbf{d}_{3}^{A}$ by the symmetry operator $C_{3z}^{2}$. We
define
\begin{equation}
E_{\mu\nu}\left(\mathbf{d}_{1}^{A}\right)\equiv\left(\begin{array}{cc}
t_{\sigma}^{(1)} & 0\\
0 & t_{\pi}^{(1)}
\end{array}\right),
\end{equation}
with the NN hopping integral defined as $t_{\sigma/\pi}^{(1)}\equiv<\phi_{x/y}\left(\mathbf{r}\right)|\hat{H}|\phi_{x/y}\left(\mathbf{r}-\mathbf{d}_{1}^{A}\right)>$.

From group theory, under the $E$-representation of $D_3$ with $p_{x,y}$-symmetric basis functions, the representation matrix for $C_{3z}$ and $C_{3z}^{2}$ can be
obtained as
\begin{equation}
D\left(C_{3z}\right)=\left(\begin{array}{cc}
-\frac{1}{2} & -\frac{\sqrt{3}}{2}\\
\frac{\sqrt{3}}{2} & -\frac{1}{2}
\end{array}\right),D\left(C_{3z}^{2}\right)=\left(\begin{array}{cc}
-\frac{1}{2} & \frac{\sqrt{3}}{2}\\
-\frac{\sqrt{3}}{2} & -\frac{1}{2}
\end{array}\right).
\end{equation}
From the $E_{\mu\nu}\left(\mathbf{d}_{1}^{A}\right)$ and by using
Eq. (3), we have
\[
E_{\mu\nu}\left(\mathbf{d}_{2}^{A}\right)=E_{\mu\nu}\left(C_{3z}\mathbf{d}_{1}^{A}\right)=D\left(C_{3z}\right)E_{\mu\nu}\left(\mathbf{d}_{1}^{A}\right)\left[D\left(C_{3z}\right)\right]^{\dagger},
\]
\begin{equation}
=\left(\begin{array}{cc}
\frac{1}{4}t_{\sigma}^{(1)}+\frac{3}{4}t_{\pi}^{(1)} & -\frac{\sqrt{3}}{4}t_{\sigma}^{(1)}+\frac{\sqrt{3}}{4}t_{\pi}^{(1)}\\
-\frac{\sqrt{3}}{4}t_{\sigma}^{(1)}+\frac{\sqrt{3}}{4}t_{\pi}^{(1)} & \frac{3}{4}t_{\sigma}^{(1)}+\frac{1}{4}t_{\pi}^{(1)}
\end{array}\right).
\end{equation}
And similarly, we have
\[
E_{\mu\nu}\left(\mathbf{d}_{3}^{A}\right)=E_{\mu\nu}\left(C_{3z}^{2}\mathbf{d}_{1}^{A}\right)=D\left(C_{3z}^{2}\right)E_{\mu\nu}\left(\mathbf{d}_{1}^{A}\right)\left[D\left(C_{3z}^{2}\right)\right]^{\dagger},
\]
\begin{equation}
=\left(\begin{array}{cc}
\frac{1}{4}t_{\sigma}^{(1)}+\frac{3}{4}t_{\pi}^{(1)} & \frac{\sqrt{3}}{4}t_{\sigma}^{(1)}-\frac{\sqrt{3}}{4}t_{\pi}^{(1)}\\
\frac{\sqrt{3}}{4}t_{\sigma}^{(1)}-\frac{\sqrt{3}}{4}t_{\pi}^{(1)} & \frac{3}{4}t_{\sigma}^{(1)}+\frac{1}{4}t_{\pi}^{(1)}
\end{array}\right).
\end{equation}

The $B$ site also has three NN $A$ sites, with the corresponding vectors $\vec{BA}$ labeled
as $\mathbf{d}_{i=1-3}^{B}$. By using the aforementioned
procedure, we can obtain $E_{\mu\nu}\left(\mathbf{d}_{i=1-3}^{B}\right)$.
In the basis $\left\{ |\phi_{x}^{A}>,|\phi_{y}^{A}>,|\phi_{x}^{B}>,|\phi_{y}^{B}>\right\} $,
the NN hopping $E_{\mu\nu}\left(\mathbf{d}_{i=1-3}^{A}\right)$ correspond
to the upper right off-diagonal $2\times2$ sub-block of the Hamiltonian,
and the bottom left off-diagonal sub-block, $E_{\mu\nu}\left(\mathbf{d}_{i=1-3}^{B}\right)$
can be more readily obtained by Hermitian conjugation. It is especially worth
emphasizing that the tight-binding model Hamiltonian obtained by symmetry here
is exactly the same as the tight-binding model Hamiltonian obtained by the Slater-Koster method in the main text. For example, we can derive $E_{\mu\nu}\left(\mathbf{d}_{2}^{A}\right)$
by using the formula Eq. (2) in the main text. As shown in Fig.\ref{Fig_TB_symmetry},
the angles in Eq. (2) in the main text read $\theta_{x,\mathbf{d}_{2}^{A}}=2\pi/3$
and $\theta_{y,\mathbf{d}_{2}^{A}}=\pi/6$ . The four matrix elements
read
\[
E_{1,1}\left(\mathbf{d}_{2}^{A}\right)=t_{\sigma}^{(1)}\cos\left(\frac{2\pi}{3}\right)^{2}+t_{\pi}^{(1)}\sin\left(\frac{2\pi}{3}\right)^{2}=\frac{1}{4}t_{\sigma}^{(1)}+\frac{3}{4}t_{\pi}^{(1)},
\]

\[
E_{1,2}\left(\mathbf{d}_{2}^{A}\right)=t_{\sigma}^{(1)}\cos\left(\frac{2\pi}{3}\right)\cos\left(\frac{\pi}{6}\right)+t_{\pi}^{(1)}\sin\left(\frac{2\pi}{3}\right)\sin\left(\frac{\pi}{6}\right)=-\frac{\sqrt{3}}{4}t_{\sigma}^{(1)}+\frac{\sqrt{3}}{4}t_{\pi}^{(1)},
\]

\[
E_{2,1}\left(\mathbf{d}_{2}^{A}\right)=t_{\sigma}^{(1)}\cos\left(\frac{\pi}{6}\right)\cos\left(\frac{2\pi}{3}\right)+t_{\pi}^{(1)}\sin\left(\frac{\pi}{6}\right)\sin\left(\frac{2\pi}{3}\right)=-\frac{\sqrt{3}}{4}t_{\sigma}^{(1)}+\frac{\sqrt{3}}{4}t_{\pi}^{(1)},
\]
\[
E_{2,2}\left(\mathbf{d}_{2}^{A}\right)=t_{\sigma}^{(1)}\cos\left(\frac{\pi}{6}\right)^{2}+t_{\pi}^{(1)}\sin\left(\frac{\pi}{6}\right)^{2}=\frac{3}{4}t_{\sigma}^{(1)}+\frac{1}{4}t_{\pi}^{(1)}.
\]
As a result, we have the same tight-binding Hamiltonian by using
the two methods.

Then let's come to the next-nearest-neighbor (NNN) hopping integrals. The $A$ site has six NNN $A$ sites, with the corresponding vectors labeled by $\mathbf{R}_{i=1-6}^{A}$, which are split into two categories
$\mathbf{R}_{i=1,3,5}^{A}$ and $\mathbf{R}_{i=4,6,2}^{A}$, as shown
in Fig.\ref{Fig_TB_symmetry}. Under the symmetry operator $C_{3z}$, $\mathbf{R}_{1}^{A}\stackrel{C_{3z}}{\rightarrow}\mathbf{R}_{3}^{A}\stackrel{C_{3z}}{\rightarrow}\mathbf{R}_{5}^{A}$
and $\mathbf{R}_{4}^{A}\stackrel{C_{3z}}{\rightarrow}\mathbf{R}_{6}^{A}\stackrel{C_{3z}}{\rightarrow}\mathbf{R}_{2}^{A}$.
We define
\begin{equation}
E_{\mu\nu}\left(\mathbf{R}_{1}^{A}\right)\equiv\left(\begin{array}{cc}
t_{\pi}^{(2)} & 0\\
0 & t_{\sigma}^{(2)}
\end{array}\right),
\end{equation}
with the NNN hopping integral defined as $t_{\pi/\sigma}^{(2)}\equiv<\phi_{x/y}\left(\mathbf{r}\right)|\hat{H}|\phi_{x/y}\left(\mathbf{r}-\mathbf{R}_{1}^{A}\right)>$.
From the $E_{\mu\nu}\left(\mathbf{R}_{1}^{A}\right)$ and by using
Eq. (3), we have
\[
E_{\mu\nu}\left(\mathbf{R}_{3}^{A}\right)=E_{\mu\nu}\left(C_{3z}\mathbf{R}_{1}^{A}\right)=D\left(C_{3z}\right)E_{\mu\nu}\left(\mathbf{R}_{1}^{A}\right)\left[D\left(C_{3z}\right)\right]^{\dagger},
\]
\begin{equation}
=\left(\begin{array}{cc}
\frac{3}{4}t_{\sigma}^{(2)}+\frac{1}{4}t_{\pi}^{(2)} & \frac{\sqrt{3}}{4}t_{\sigma}^{(2)}-\frac{\sqrt{3}}{4}t_{\pi}^{(2)}\\
\frac{\sqrt{3}}{4}t_{\sigma}^{(2)}-\frac{\sqrt{3}}{4}t_{\pi}^{(2)} & \frac{1}{4}t_{\sigma}^{(2)}+\frac{3}{4}t_{\pi}^{(2)}
\end{array}\right).
\end{equation}
And
\[
E_{\mu\nu}\left(\mathbf{R}_{5}^{A}\right)=E_{\mu\nu}\left(C_{3z}^{2}\mathbf{R}_{1}^{A}\right)=D\left(C_{3z}^{2}\right)E_{\mu\nu}\left(\mathbf{R}_{1}^{A}\right)\left[D\left(C_{3z}^{2}\right)\right]^{\dagger},
\]
\begin{equation}
=\left(\begin{array}{cc}
\frac{3}{4}t_{\sigma}^{(2)}+\frac{1}{4}t_{\pi}^{(2)} & -\frac{\sqrt{3}}{4}t_{\sigma}^{(2)}+\frac{\sqrt{3}}{4}t_{\pi}^{(2)}\\
-\frac{\sqrt{3}}{4}t_{\sigma}^{(2)}+\frac{\sqrt{3}}{4}t_{\pi}^{(2)} & \frac{1}{4}t_{\sigma}^{(2)}+\frac{3}{4}t_{\pi}^{(2)}
\end{array}\right).
\end{equation}

As aforementioned, the emergent honeycomb lattice approximately has $D_{6}$ point-group symmetry. $\mathbf{R}_{1}^{A}$ is rotated to $\mathbf{R}_{4}^{A}$ by $C_{2z}=C_{6z}^3$, whose representation matrix is $-I$ with $I$ being a unit matrix.  Thus we obtain $E_{\mu\nu}\left(\mathbf{R}_{4}^{A}\right)=E_{\mu\nu}\left(\mathbf{R}_{1}^{A}\right),$
and we can directly write $E_{\mu\nu}\left(\mathbf{R}_{6}^{A}\right)=E_{\mu\nu}\left(\mathbf{R}_{3}^{A}\right)$
and $E_{\mu\nu}\left(\mathbf{R}_{2}^{A}\right)=E_{\mu\nu}\left(\mathbf{R}_{5}^{A}\right)$
. The $B$ site also has six NNN $B$ sites, the corresponding vectors labeled by $\mathbf{R}_{i=1-6}^{B}$,
and $\mathbf{R}_{i=1-6}^{B}$ = $\mathbf{R}_{i=1-6}^{A}$, therefore,
$E_{\mu\nu}\left(\mathbf{R}_{i=1-6}^{B}\right)=E_{\mu\nu}\left(\mathbf{R}_{i=1-6}^{A}\right)$.
And we can check that the NNN tight-binding Hamiltonian by the two
methods are also exactly the same. Consequently, we have re-derived the same total tight-binding
Hamiltonian here by strict and precise symmetry method, which justifies
our tight-binding model in the main text.

\section{Appendix:The explicit formulism of our Hamiltonian in the reciprocal space and the robust nesting}
In the basis $\left\{ p_{x}^{A},p_{y}^{A},p_{x}^{B},p_{y}^{B}\right\} $,
where A, B denote the sublattice in the emergent honeycomb lattice
superstructure, the total Hamiltonian up to the next nearest neighbor hopping terms in the reciprocal space can be written as
\begin{equation}
H\left(k\right)=H_{0}+H_{1}+H_{2},
\end{equation}
where the three terms are the chemical potential, the nearest neighbor hopping and the next nearest neighbor hopping terms respectively.
\begin{equation}
H_{0}=-\mu I_{4},
\end{equation}
\begin{equation}
H_{1}=\left(\begin{array}{cccc}
0 & 0 & h13 & h14\\
 & 0 & h23 & h24\\
 & \dagger & 0 & 0\\
 &  &  & 0
\end{array}\right),
\end{equation}
\begin{equation}
H_{2}=\left(\begin{array}{cccc}
hn11 & hn12 & 0 & 0\\
 & hn22 & 0 & 0\\
 & \dagger & hn11 & hn12\\
 &  &  & hn22
\end{array}\right).
\end{equation}

The above matrix elements are given as
\begin{equation}
\begin{split}
&h13=\frac{1}{4}(t_{\pi}+3t_{\sigma})\left(e^{i\left(\frac{k_{y}}{2\sqrt{3}}-\frac{k_{x}}{2}\right)}+e^{i\left(\frac{k_{y}}{2\sqrt{3}}+\frac{k_{x}}{2}\right)}\right)+t_{\pi}e^{-i\frac{k_{y}}{\sqrt{3}}}, \\
& h14=-\frac{\sqrt{3}}{4}(t_{\pi}-t_{\sigma})\left(-1+e^{ik_{x}}\right)e^{-\frac{1}{6}i\left(3k_{x}-\sqrt{3}k_{y}\right)},\\
&h23=h14,\\
&h24=\frac{1}{4}(3t_{\pi}+t_{\sigma})\left(e^{i\left(\frac{k_{y}}{2\sqrt{3}}-\frac{k_{x}}{2}\right)}+e^{i\left(\frac{k_{y}}{2\sqrt{3}}+\frac{k_{x}}{2}\right)}\right)+t_{\sigma}e^{-i\frac{k_{y}}{\sqrt{3}}},\\
&hn11=(3t_{\pi2}+t_{\sigma2})\cos\left(\frac{k_{x}}{2}\right)\cos\left(\frac{\sqrt{3}k_{y}}{2}\right)+2t_{\sigma2}\cos\left(k_{x}\right),\\
&hn12=\sqrt{3}(t_{\pi2}-t_{\sigma2})\sin\left(\frac{k_{x}}{2}\right)\sin\left(\frac{\sqrt{3}k_{y}}{2}\right),\\
&hn22=(t_{\pi2}+3t_{\sigma2})\cos\left(\frac{k_{x}}{2}\right)\cos\left(\frac{\sqrt{3}k_{y}}{2}\right)+2t_{\pi2}\cos\left(k_{x}\right),
\end{split}
\end{equation}
where $t_{\pi},t_{\sigma} (t_{\pi2},t_{\sigma2})$ are the (next) nearest hopping Slater-Koster parameters.

In addition to the Slater-Koster parameters used in the text, we also choose  two other sets of  parameters to address the robustness of FS nesting at the Van-Hove dopings, with one set of Slater-Koster parameters chosen as $t^{(1)}_{\sigma}=2$ meV, $t^{(1)}_{\pi}=t^{(1)}_{\sigma}/1.56$, $t^{(2)}_{\sigma}=t^{(1)}_{\sigma}/10$, and $t^{(2)}_{\pi}=t^{(2)}_{\sigma}/1.56$
in Fig.\ref{nesting_robust} (a)(b), and the other set chosen as $t^{(1)}_{\sigma}=2$ meV, $t^{(1)}_{\pi}=t^{(1)}_{\sigma}/2$, $t^{(2)}_{\sigma}=t^{(1)}_{\sigma}/7$, and $t^{(2)}_{\pi}=t^{(2)}_{\sigma}/2$ in Fig.\ref{nesting_robust} (c)(d). All the three cases clearly show the nearly perfect FS-nesting, and thus demonstrating that the FS nesting is robust. Furthermore, the VH dopings for the three groups of parameters are all near $\pm\frac{1}{2}$. Note that due to the relationship $t^{(1)}_{\sigma}/t^{(1)}_{\pi}=t^{(2)}_{\sigma}/t^{(2)}_{\pi}$ adopted here, the absolute values of $\delta_V$ at the negative and positive doping parts are approximately equal. Such equivalence relation will be slightly changed if we do not adopt the relationship. It should be noticed that the peak in DOS (Fig.1 (c) in text) below the red dashed line come from the flat band bottom, where the FSs shown in Fig.\ref{FS_Gamma} only contain small pockets. No VHS or FS-nesting can be found there.

\begin{figure}
\scalebox{0.4}{{\includegraphics[scale=0.6]{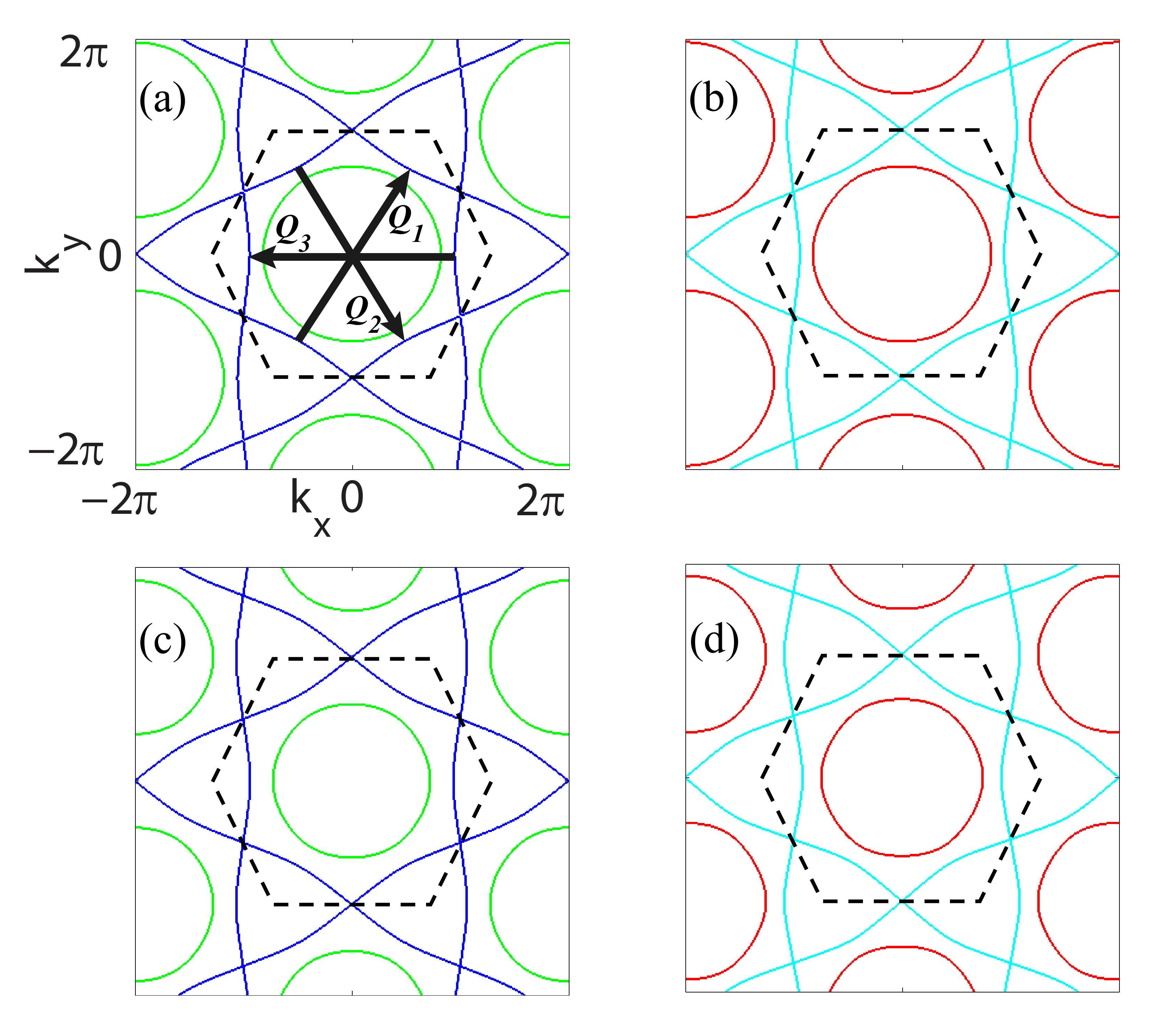}}}
\caption{The FS nesting and the Lifshitz transition for the other two parameter sets. (a)(b) FS nesting and the Lifshitz transition at doping -0.43 and 0.43 for one set of parameters. (c)(d) FS nesting and the Lifshitz transition for the other set of parameters at doping -0.48 and 0.48. The central hexagons in every plot in black dashed line indicate the Brillioun zone. The three arrows $Q_{1,2,3}$ in (a) label the three FS nesting vectors, and are omitted in (b)-(d).} \label{nesting_robust}
\end{figure}

\begin{figure}
\scalebox{0.4}{{\includegraphics[scale=0.75]{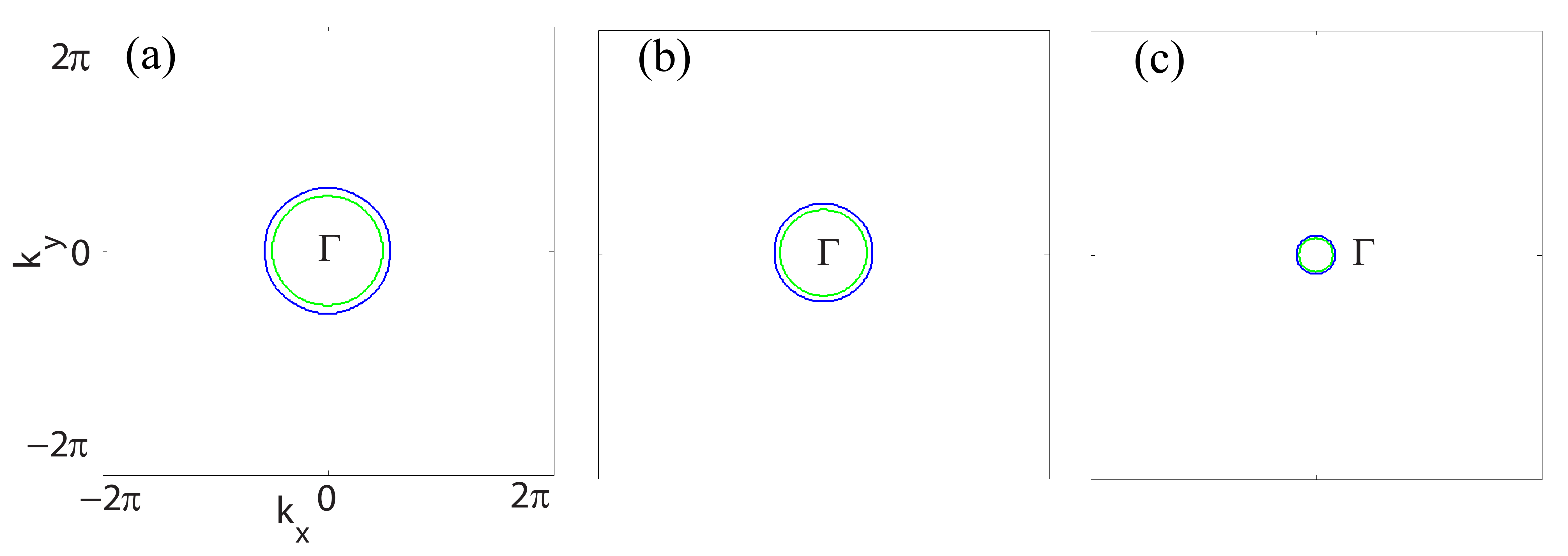}}}
\caption{The FS for the basin-like bandstructure with a quite flat base, which results in  a remarkable peak in DOS (Fig.1 (c) in text) below the red dashed line. (a)(b)(c) Two electron pockets FS around the $\Gamma$ point in the first BZ are plotted with doping -0.81, -0.88 and -0.98, respectively.} \label{FS_Gamma}
\end{figure}

\section{Appendix:The multi-orbital RPA approach}
It's proposed here that the SC detected in the MA-TBG is not driven by electron-phonon coupling. The reason behind this point lies in that only acoustic phonon modes with wave length comparable or longer than the size of a unit cell in the Moire pattern can efficiently couple with our low energy effective orbitals through changing the hopping integral shown in Eq.(2) in the main text.  The Debye frequency of such phonon modes will be too low to support the "high $T_c$ SC". The SC detected here is more likely to be driven by electron-electron interactions. We adopt the following repulsive Hubbard-Hund model listed in Eq.(3) in the main text,

The Hamiltonian adopted in our calculations is
\begin{align}
H&=H_{tb}+H_{int}\nonumber\\
&H_{int}=U\sum_{i\mu}n_{i\mu\uparrow}n_{i\mu\downarrow}+
V\sum_{i}n_{ix}n_{iy}+J_{H}\sum_{i}\Big[
\sum_{\sigma\sigma^{\prime}}c^{\dagger}_{ix\sigma}c^{\dagger}_{iy\sigma^{\prime}}
c_{ix\sigma^{\prime}}c_{iy\sigma}+(c^{\dagger}_{ix\uparrow}c^{\dagger}_{ix\downarrow}
c_{iy\downarrow}c_{iy\uparrow}+h.c.)\Big]\label{H-H-model_app}
\end{align}
where $i$ is site index, $\mu$ is orbital index, $x$ and $y$ denote orbitals $p_x$ and $p_y$, respectively, $\sigma$ and $\sigma^{\prime}$ are spin indices.

Let's define the following bare susceptibility for the non-interacting case ($U=V=J_H=0$),
\begin{align}
\chi^{(0)l_{1},l_{2}}_{l_{3},l_{4}}\left(\bm{q},\tau\right)\equiv
\frac{1}{N}\sum_{\bm{k}_{1},\bm{k}_{2}}\left<T_{\tau}c^{\dagger}_{l_{1}}(\bm{k}_{1},\tau)
c_{l_{2}}(\bm{k}_{1}+\bm{q},\tau)c^{\dagger}_{l_{3}}(\bm{k}_{2}+\bm{q},0)c_{l_{4}}(\bm{k}_{2},0)
\right>_0,\label{bare}
\end{align}
where $l_{i}$ $(i=1,\cdots,4)$ denote orbital indices. The explicit formulism of $\chi^{(0)}$ in the momentum-frequency space is,
\begin{align}
\chi^{(0)l_{1},l_{2}}_{l_{3},l_{4}}\left(\bm{q},i\omega_n\right)
=\frac{1}{N}\sum_{\bm{k},\alpha,\beta}
\xi^{\alpha}_{l_{4}}(\bm{k})\xi_{l_{1}}^{\alpha,*}(\bm{k})
\xi^{\beta}_{l_{2}}(\bm{k}+\bm{q})\xi^{\beta,*}_{l_{3}}(\bm{k}+\bm{q})
\frac{n_{F}(\varepsilon^{\beta}_{\bm{k+q}})-n_{F}
(\varepsilon^{\alpha}_{\bm{k}})}{i\omega_n
+\varepsilon^{\alpha}_{\bm{k}}-\varepsilon^{\beta}_{\bm{k}+\bm{q}}},\label{explicit_free}
\end{align}
where $\alpha,\beta=1,...,4$ are band indices, $\varepsilon^{\alpha}_{\bm{k}}$ and $\xi^{\alpha}_{l}\left(\bm{k}\right)$ are the $\alpha-$th eigenvalue and eigenvector
of the $h(\bm{k})$ matrix respectively and $n_F$ is the Fermi-Dirac distribution
function.
\begin{figure}[tb]
\centerline{\includegraphics[height=10 cm]{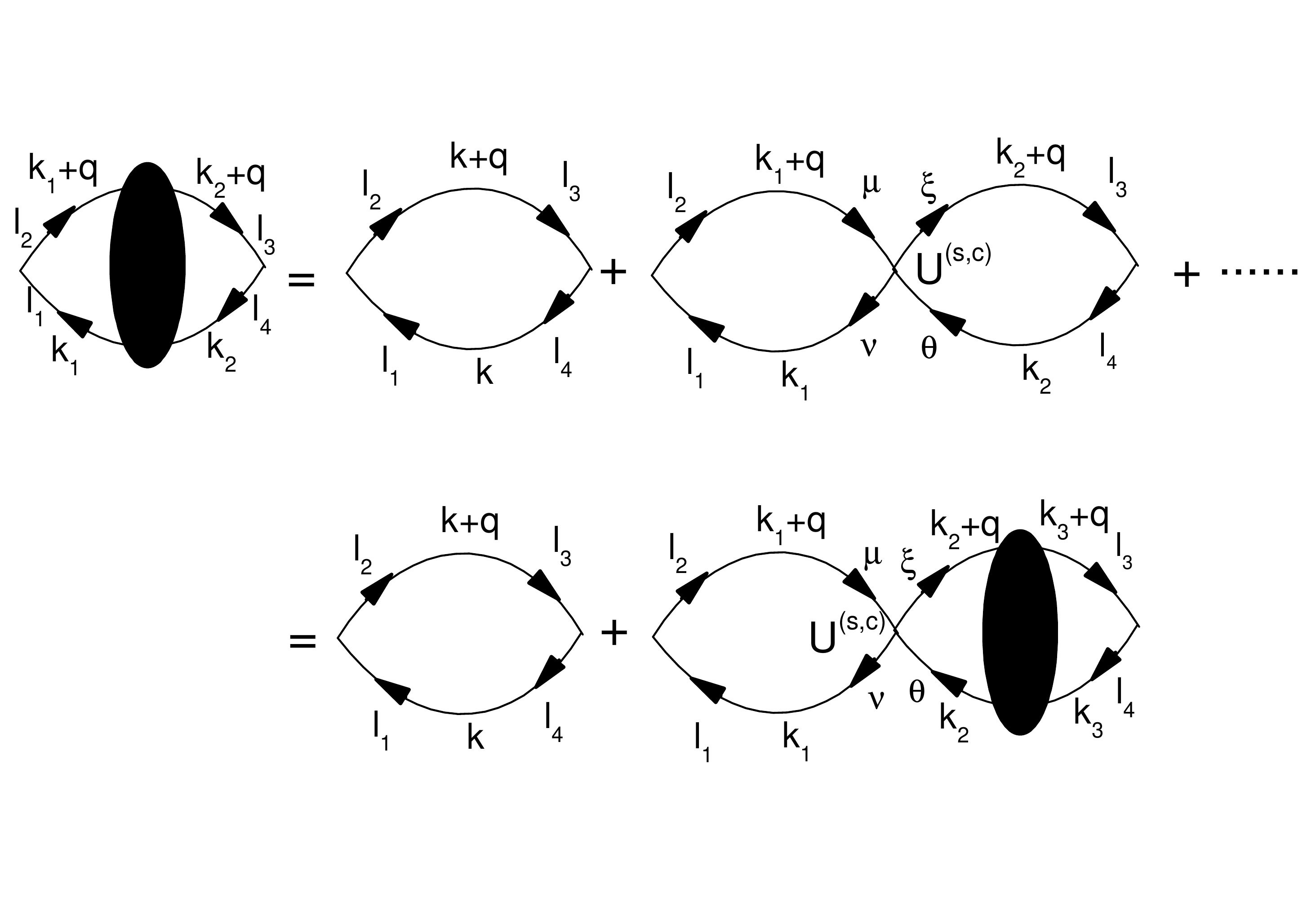}}
\caption{(color online). Feyman's diagram for the renormalized susceptibilities in the RPA level.  \label{RPA_diagram} }
\end{figure}

When interactions turn on, we define the spin ($\chi^{(s)}$) and charge ($\chi^{(c)}$) susceptibility as follow,
\begin{align}
\chi^{(c)l_{1},l_{2}}_{l_{3},l_{4}}\left(\bm{q},\tau\right)\equiv&
\frac{1}{2N}\sum_{\bm{k}_{1},\bm{k}_{2},\sigma_{1},\sigma_{2}}\left<T_{\tau}
C^{\dagger}_{l_{1},\sigma_{1}}(\bm{k}_{1},\tau)
C_{l_{2},\sigma_{1}}(\bm{k}_{1}+\bm{q},\tau)
C^{\dagger}_{l_{3},\sigma_{2}}(\bm{k}_{2}+\bm{q},0)C_{l_{4},\sigma_{2}}
(\bm{k}_{2},0)\right>,
\nonumber\\
\chi^{\left(s^{z}\right)l_{1},l_{2}}_{l_{3},l_{4}}
\left(\bm{q},\tau\right)\equiv&
\frac{1}{2N}\sum_{\bm{k}_{1},\bm{k}_{2},\sigma_{1},\sigma_{2}}\sigma_{1}\sigma_{2}
\left<T_{\tau}C^{\dagger}_{l_{1},\sigma_{1}}(\bm{k}_{1},\tau)
C_{l_{2},\sigma_{1}}(\bm{k}_{1}+\bm{q},\tau)C^{\dagger}_{l_{3},\sigma_{2}}
(\bm{k}_{2}+\bm{q},0)
C_{l_{4},\sigma_{2}}(\bm{k}_{2},0)\right>,\nonumber\\
\chi^{\left(s^{+-}\right)l_{1},l_{2}}_{l_{3},l_{4}}\left(\bm{q},\tau\right)
\equiv&
\frac{1}{N}\sum_{\bm{k}_{1},\bm{k}_{2}}\left<T_{\tau}
C^{\dagger}_{l_{1}\uparrow}(\bm{k}_{1},\tau)
C_{l_{2}\downarrow}(\bm{k}_{1}+\bm{q},\tau)C^{\dagger}_{l_{3}\downarrow}
(\bm{k}_{2}+\bm{q},0)
C_{l_{4}\uparrow}(\bm{k}_{2},0)\right>,\nonumber\\
\chi^{\left(s^{-+}\right)l_{1},l_{2}}_{l_{3},l_{4}}\left(\bm{q},\tau\right)\equiv&
\frac{1}{N}\sum_{\bm{k}_{1},\bm{k}_{2}}\left<T_{\tau}
C^{\dagger}_{l_{1}\downarrow}(\bm{k}_{1},\tau)
C_{l_{2}\uparrow}(\bm{k}_{1}+\bm{q},\tau)C^{\dagger}_{l_{3}\uparrow}
(\bm{k}_{2}+\bm{q},0)
C_{l_{4}\downarrow}(\bm{k}_{2},0)\right>.
\end{align}
Note that in non-magnetic state we have
$\chi^{\left(s^{z}\right)}=\chi^{\left(s^{+-}\right)}=\chi^{\left(s^{-+}\right)}\equiv\chi^{\left(s\right)}$, and when $U=V=J_H=0$ we have $\chi^{(c)}=\chi^{(s)}=\chi^{(0)}$.
In the RPA level, the renormalized spin/charge susceptibilities for the system are,
\begin{align}
\chi^{(s)}\left(\bm{q},i\nu\right)=&\left[I-\chi^{(0)}
\left(\bm{q},i\nu\right)U^{(s)}\right]^{-1}\chi^{(0)}
\left(\bm{q},i\nu\right),\nonumber\\
\chi^{(c)}\left(\bm{q},i\nu\right)=&\left[I+\chi^{(0)}
\left(\bm{q},i\nu\right)U^{(c)}\right]^{-1}\chi^{(0)}
\left(\bm{q},i\nu\right),\label{RPA_SUS}
\end{align}
where $\chi^{(s,c)}\left(\bm{q},i\nu_{n}\right)$, $\chi^{(0)}\left(\bm{q},i\nu_{n}\right)$ and $U^{(s,c)}$ are operated as
$16\times 16$ matrices (the upper or lower two indices are viewed as one number). Labelling orbitals $\left\{ p_{x}^{A},p_{y}^{A},p_{x}^{B},p_{y}^{B}\right\} $ as $\left\{ 1,2,3,4\right\} $, the nonzero elements of the matrix $U^{(s,c)l_{1}l_{2}}_{l_{3}l_{4}}$ are listed as follows:
\begin{align}
&U^{(s)11}_{11}=U^{(s)22}_{22}=U^{(s)33}_{33}=U^{(s)44}_{44}=U           \nonumber\\
&U^{(s)11}_{22}=U^{(s)22}_{11}=U^{(s)33}_{44}=U^{(s)44}_{33}=J_{H}       \nonumber\\
&U^{(s)12}_{12}=U^{(s)21}_{21}=U^{(s)34}_{34}=U^{(s)43}_{43}=J_{H}       \nonumber\\
&U^{(s)12}_{21}=U^{(s)21}_{12}=U^{(s)34}_{43}=U^{(s)43}_{34}=V
\end{align}

\begin{align}
&U^{(c)11}_{11}=U^{(c)22}_{22}=U^{(c)33}_{33}=U^{(c)44}_{44}=U           \nonumber\\
&U^{(c)11}_{22}=U^{(c)22}_{11}=U^{(c)33}_{44}=U^{(c)44}_{33}=2V-J_{H}    \nonumber\\
&U^{(c)12}_{12}=U^{(c)21}_{21}=U^{(c)34}_{34}=U^{(c)43}_{43}=J_{H}       \nonumber\\
&U^{(c)12}_{21}=U^{(c)21}_{12}=U^{(c)34}_{43}=U^{(c)43}_{34}=2J_{H}-V
\end{align}
The Feyman's diagram of RPA is shown in Fig.\ref{RPA_diagram} For repulsive Hubbard-interactions, the spin susceptibility is enhanced and the charge susceptibility is suppressed. Note that there is a critical interaction strength $U_c$ which depends on the ratio $J_H/U$. When $U>U_c$, the denominator
matrix $I-\chi^{(0)}\left(\bm{q},i\nu\right)U^{(s)}$ in Eq.(\ref{RPA_SUS}) will have zero eigenvalues for some $\bm{q}$ and the renormalized spin susceptibility diverges there, which invalidates the RPA treatment. This divergence of spin susceptibility for $U>U_c$ implies magnetic order. When $U<U_c$, the short-ranged spin or charge fluctuations would mediate Cooper pairing in the system.

\begin{figure}
\scalebox{0.4}{{\includegraphics[scale=0.55]{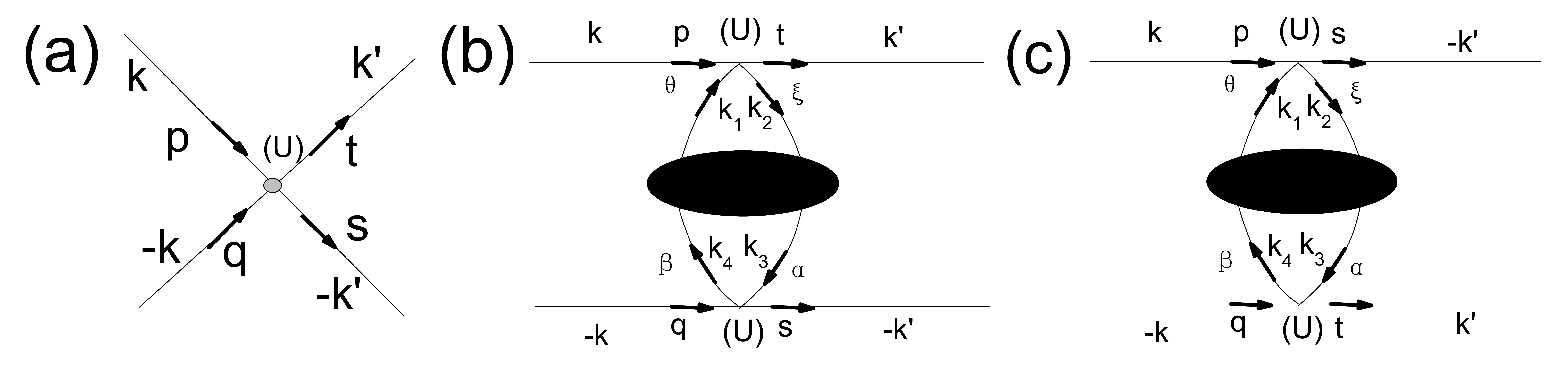}}}
\caption{The three processes which contribute the renormalized effective vertex considered in the RPA, with (a) the bare interation vertex and (b) (c) the two second order perturbation processes during which spin or charge fluctuations are exchanged between a cooper pair.} \label{RPA_effective}
\end{figure}

Let's consider a Cooper pair with momentum/orbital $(\bm{k'}t,-\bm{k'}s)$, which could be scattered to
$(\bm{k}p,-\bm{k}q)$ by exchanging charge or spin fluctuations. In the RPA level, The effective interaction induced by this process is as follow,
\begin{align}
V^{RPA}_{eff}=\frac{1}{N}\sum_{pqst,\bm{k}\bm{k}'}\Gamma^{pq}_{st}(\bm{k},\bm{k}')
c_{p}^{\dagger}(\bm{k})c_{q}^{\dagger}(-\bm{k})c_{s}(-\bm{k}')c_{t}(\bm{k}'),
\label{effective_vertex}
\end{align}
We consider the three processes in Fig.\ref{RPA_effective} which contribute to the effective vertex $\Gamma^{pq}_{st}(\bm{k},\bm{k}')$, where (a) represents the bare interaction vertex, (b) and (c) represent the two second order perturbation processes during which spin or charge fluctuations are exchanged between a cooper pair. In the singlet channel, the effective vertex $\Gamma^{pq}_{st}(\bm{k},\bm{k}')$ is given as follow,
\begin{align}
\Gamma^{pq(s)}_{st}(\bm{k},\bm{k}')=\left(\frac{U^{(c)}+3U^{(s)}}{4}\right)^{pt}_{qs}+
\frac{1}{4}\left[3U^{(s)}\chi^{(s)}\left(\bm{k}-\bm{k}'\right)U^{(s)}-U^{(c)}
\chi^{(c)}\left(\bm{k}-\bm{k}'\right)U^{(c)}\right]^{pt}_{qs}+\nonumber\\
\frac{1}{4}\left[3U^{(s)}\chi^{(s)}\left(\bm{k}+\bm{k}'\right)U^{(s)}-U^{(c)}
\chi^{(c)}\left(\bm{k}+\bm{k}'\right)U^{(c)}\right]^{ps}_{qt},
\end{align}
while in the triplet channel, it is
\begin{align}
\Gamma^{pq(t)}_{st}(\bm{k},\bm{k}')=\left(\frac{U^{(c)}-U^{(s)}}{4}\right)^{pt}_{qs}-
\frac{1}{4}\left[U^{(s)}\chi^{(s)}\left(\bm{k}-\bm{k}'\right)U^{(s)}+U^{(c)}
\chi^{(c)}\left(\bm{k}-\bm{k}'\right)U^{(c)}\right]^{pt}_{qs}+\nonumber\\
\frac{1}{4}\left[U^{(s)}\chi^{(s)}\left(\bm{k}+\bm{k}'\right)U^{(s)}+U^{(c)}
\chi^{(c)}\left(\bm{k}+\bm{k}'\right)U^{(c)}\right]^{ps}_{qt},
\end{align}
Notice that the vertex $\Gamma^{pq}_{st}(\bm{k},\bm{k}')$ has been symmetrized for the singlet case and anti-symmetrized for the
triplet case. Generally we neglect the frequency-dependence of $\Gamma$ and replace it by $\Gamma^{pq}_{st}(\bm{k},\bm{k}')\approx\Gamma^{pq}_{st}(\bm{k},\bm{k}',0)$. Usually, only the real part of $\Gamma$ is adopted \cite{Scalapino1,Scalapino2}

Considering only intra-band pairings, we obtain the following effective pairing interaction on the FS,
\begin{align}
V_{eff}=
\frac{1}{N}\sum_{\alpha\beta,\bm{k}\bm{k'}}V^{\alpha\beta}(\bm{k,k'})c_{\alpha}^{\dagger}(\bm{k})
c_{\alpha}^{\dagger}(-\bm{k})c_{\beta}(-\bm{k}')c_{\beta}(\bm{k}'),\label{pairing_interaction}
\end{align}
where $\alpha,\beta=1,\cdots,4$ are band indices and $V^{\alpha\beta}(\bm{k,k'})$ is
\begin{align}
V^{\alpha\beta}(\bm{k,k'})=\sum_{pqst,\bm{k}\bm{k'}}\Gamma^{pq}_{st}(\bm{k,k'},0)\xi_{p}^{\alpha,*}(\bm{k})
\xi_{q}^{\alpha,*}(-\bm{k})\xi_{s}^{\beta}(-\bm{k'})\xi_{t}^{\beta}(\bm{k'}).\label{effective_potential}
\end{align}
From the effective pairing interaction (\ref{pairing_interaction}), one can obtain the following linearized gap
equation \cite{Scalapino1,Scalapino2} to determine the $T_c$ and the leading pairing symmetry of the system,
\begin{align}
-\frac{1}{(2\pi)^2}\sum_{\beta}\oint_{FS}
dk'_{\Vert}\frac{V^{\alpha\beta}(\bm{k,k'})}{v^{\beta}_{F}(\bm{k'})}
\Delta_{\beta}(\bm{k'})=\lambda
\Delta_{\alpha}(\bm{k}).\label{eigenvalue_Tc}
\end{align}
This equation can be looked upon as an eigenvalue problem, where the normalized eigenvector $\Delta_{\alpha}(\bm{k})$ represents the relative gap function on the $\alpha-$th FS patches near $T_c$, and the eigenvalue $\lambda$ determines $T_c$ via $T_{c}=E_ce^{-1/\lambda}$, where the cut off energy $E_c$ is of order band-width. The leading pairing symmetry is determined by the largest eigenvalue $\lambda$ of Eq. (\ref{eigenvalue_Tc}).

\section{Appendix:Classification of the pairing symmetry}

Let's start from the effective Hamiltonian obtained from exchanging spin fluctuations:
\begin{align}
H=H_{tb}+\frac{1}{N}\sum_{\bm{k}\bm{k}'\alpha\beta}
V^{\alpha\beta}(\bm{k},\bm{k}')
c^{\dagger}_{\alpha}(\bm{k})
c^{\dagger}_{\alpha}(-\bm{k})
c_{\beta}(-\bm{k}')
c_{\beta}(\bm{k}')
\end{align}
This normal state Hamiltonian should be invariant under the point group $G\equiv\{\hat{P}\}$, where $\hat{P}$ is any operation in the point group. The action of $\hat{P}$ on any electron operator is
\begin{eqnarray}
\hat{P}c_{\alpha}(\bm{k})\hat{P}^{\dagger}=c_{\alpha}(\hat{P}\bm{k}).
\end{eqnarray}
Note that the band index $\alpha$ will not be changed by symmetry operation. From the invariant of $H$ under the point group $G$, i.e.  $\hat{P}H\hat{P}^{\dagger}=H$, we have
\begin{eqnarray}\label{symmetry}
V^{\alpha\beta}(\hat{P}\bm{k},\hat{P}\bm{k}')
=V^{\alpha\beta}(\bm{k},\bm{k}')
\end{eqnarray}

The linearized gap equation
\begin{align}
-\frac{1}{(2\pi)^2}\sum_{\beta}\oint_{FS}
dk'_{\Vert}\frac{V^{\alpha\beta}(\bm{k,k'})}{v^{\beta}_{F}(\bm{k'})}
\Delta_{\beta}(\bm{k'})=\lambda
\Delta_{\alpha}(\bm{k}).\label{eigenvalue_Tc2}
\end{align}
can be rewritten as
\begin{align}
\frac{1}{(2\pi)^2}\sum_{\beta}\iint_{\Delta E}
d^{2}\bm{k}'V^{\alpha\beta}(\bm{k,k'})
\Delta_{\beta}(\bm{k'})=-\lambda \Delta E
\Delta_{\alpha}(\bm{k}).\label{eigenvalue_Tc3}
\end{align}
where the integral $\iint$ is performed within a narrow energy window near the FS with the width of the window $\Delta E\to 0$. After discreteness in the lattice, Eq. (\ref{eigenvalue_Tc3}) can be taken as an eigenvalue problem with $\lambda$ proportional to the eigenvalue and $\Delta_{\alpha}(\bm{k})$ proportional to the eigenvector.

From Eq. (\ref{eigenvalue_Tc3}) and Eq. (\ref{symmetry}), we can find that each solved $\Delta_{\alpha}(\bm{k})$ belong to an irreducible representation of the point group. In Table I, we list all the irreducible representation and the basis functions of the $D_6$ point group of our model in two spatial dimensions, with the pairing symmetry of each basic function marked. There are 5 different pairing symmetries, i.e. s-wave, $(p_x,p_y)$ (degenerate $p$-wave), $(d_{x^2-y^2},d_{xy})$ (degenerate $d$-wave), $f_{x\left(x^{2}-3y^{2}\right)}$ ($f$-wave), $f_{y\left(3x^{2}-y^{2}\right)}$ ($f'$-wave), each has definite parity, either even or odd.

The real material has the $D_3$ point group. In Table II, we list all its irreducible representation and its basis functions in two spatial dimensions. In general, the irreducible representation, and hence the classification of the pairing symmetry for $D_6$- and $D_3$- point-groups are different. For example, the $s$-wave and $f$-wave belong to different irreducible representations for $D_6$ point-group and thus they generally will not mix, but they belong to the same irreducible representations for $D_3$ point-group and thus will generally mix. However, in the presence with SU(2) spin symmetry (i.e. without SOC), the Pauli's principle requires the gap function to be either odd or even according to whether the spin status is triplet or singlet. In this case, the odd function and even function in the same irreducible representation of $D_3$ can be distinguished through spin status and generally will not mix, and therefore the two point groups have the same pairing symmetry classification, i.e., $s, p, d, f, f'$.

\begin{table}
\caption{Character table for point group $D_{6}$ and possible superconductivity
pairing symmetry.}
\begin{tabular}{|c|c|c|c|c|c|c|c|c|}
\hline
$D_{6}$ & $E$ & $2C_{6}\left(z\right)$ & $2C_{3}\left(z\right)$ & $C_{2}\left(z\right)$ & $3C_{2}^{'}$ & $3C_{2}^{''}$ & odd functions & even functions\tabularnewline
\hline
$A_{1}$ & $+1$ & $+1$ & $+1$ & $+1$ & $+1$ & $+1$ & \textemdash{} & $\left(x^{2}+y^{2}\right)$ $s$-wave\tabularnewline
\hline
$A_{2}$ & $+1$ & $+1$ & $+1$ & $+1$ & $-1$ & $-1$ & \textemdash{} & \textemdash{}\tabularnewline
\hline
$B_{1}$ & $+1$ & $-1$ & $+1$ & $-1$ & $+1$ & $-1$ & $x\left(x^{2}-3y^{2}\right)$ $f$-wave & \textemdash{}\tabularnewline
\hline
$B_{2}$ & $+1$ & $-1$ & $+1$ & $-1$ & $-1$ & $+1$ & $y\left(3x^{2}-y^{2}\right)$ $f'$-wave & \textemdash{}\tabularnewline
\hline
$E_{1}$ & $+2$ & $+1$ & $-1$ & $-2$ & 0 & 0 & $\left(x,y\right)$ $p$-wave & \textemdash{}\tabularnewline
\hline
$E_{2}$ & $+2$ & $-1$ & $-1$ & $+2$ & 0 & $0$ & \textemdash{} & $\left(x^{2}-y^{2},xy\right)$ $d$-wave\tabularnewline
\hline
\end{tabular}
\end{table}

\begin{table}
\caption{Character table for point group $D_{3}$ in case with spin-$SU(2)$ symmetry and possible superconductivity
pairing symmetry.}
\begin{tabular}{|c|c|c|c|c|c|}
\hline
$D_{3}$ & $E$ & $2C_{3}\left(z\right)$ & $3C_{2}^{'}$ & odd functions & even functions\tabularnewline
\hline
$A_{1}$ & $+1$ & $+1$ & $+1$ & $x\left(x^{2}-3y^{2}\right)$ $f$-wave & $\left(x^{2}+y^{2}\right)$ $s$-wave\tabularnewline
\hline
$A_{2}$ & $+1$ & $+1$ & $-1$ & $y\left(3x^{2}-y^{2}\right)$ $f'$-wave & \textemdash{}\tabularnewline
\hline
$E$ & $+2$ & $-1$ & $0$ & $\left(x,y\right)$
$p$-wave & $\left(x^{2}-y^{2},xy\right)$ $d$-wave\tabularnewline
\hline
\end{tabular}
\end{table}

\section{Appendix:The $d+id$ superconducting state}

Since the $d_{x^2-y^2}$ and $d_{xy}$ pairing states are degenerate, they would probably mix to lower the energy below the critical temperature $T_c$. If we consider the Hamiltonian with the interaction between the spin $\uparrow$ and $\downarrow$:
\begin{align}
H=\sum_{\bm{k}\alpha\sigma}\varepsilon^{\alpha}_{\bm{k}}
c^{\dagger}_{\alpha\sigma}(\bm{k})c_{\alpha\sigma}(\bm{k})
+\frac{1}{N}\sum_{\bm{k}\bm{k}'\alpha\beta}V^{\alpha\beta}(\bm{k},\bm{k}')
c^{\dagger}_{\alpha\uparrow}(\bm{k})c^{\dagger}_{\alpha\downarrow}(-\bm{k})
c_{\beta\downarrow}(-\bm{k}')c_{\beta\uparrow}(\bm{k}')
\end{align}
the total mean-field energy
\begin{align}\label{ene}
E=&\langle H\rangle
=\sum_{\bm{k}\alpha\sigma}\varepsilon^{\alpha}_{\bm{k}}
\Big\langle c^{\dagger}_{\alpha\sigma}(\bm{k})
c_{\alpha\sigma}(\bm{k})\Big\rangle
+\frac{1}{N}\sum_{\bm{k}\bm{k}'\alpha\beta}
V^{\alpha\beta}(\bm{k},\bm{k}')
\Big\langle c^{\dagger}_{\alpha\uparrow}(\bm{k})
c^{\dagger}_{\alpha\downarrow}(-\bm{k})\Big\rangle
\Big\langle c_{\beta\downarrow}(-\bm{k}')
c_{\beta\uparrow}(\bm{k}')\Big\rangle                                     \nonumber\\
=&\sum_{\bm{k}\alpha}\varepsilon^{\alpha}_{\bm{k}}
\left[1-\frac{\varepsilon^{\alpha}_{\bm{k}}-\mu}
{\sqrt{(\varepsilon^{\alpha}_{\bm{k}}-\mu)^2
+|\Delta^{\alpha}_{\bm{k}}|^2}}\right]+\frac{1}{4N}\sum_{\bm{k}\bm{k}'\alpha\beta}
V^{\alpha\beta}(\bm{k},\bm{k}')\frac{(\Delta^{\alpha}_{\bm{k}})^*}
{\sqrt{(\varepsilon^{\alpha}_{\bm{k}}-\mu)^2
+|\Delta^{\alpha}_{\bm{k}}|^2}}
\frac{\Delta^{\beta}_{\bm{k}'}}
{\sqrt{(\varepsilon^{\beta}_{\bm{k}'}-\mu)^2
+|\Delta^{\beta}_{\bm{k}'}|^2}},
\end{align}
where the chemical potential $\mu$ is determined by the constraint of the average electron number in the superconducting state. If we further set $\Delta^{\alpha}_{\bm{k}}\equiv\frac{1}{N}\sum_{\bm{k}'\beta}
V^{\alpha\beta}(\bm{k},\bm{k}')\Big\langle c_{\beta\downarrow}(-\bm{k}')
c_{\beta\uparrow}(\bm{k}')\Big\rangle=K_1d^{\alpha}_{x^2-y^2}(\bm{k})+(K_2+iK_3)
d^{\alpha}_{xy}(\bm{k})$, where $d^{\alpha}_{x^2-y^2}(\bm{k})$ and $d^{\alpha}_{xy}(\bm{k})$ denote the normalized gap functions of corresponding symmetries, the mixing coefficients $K_1$, $K_2$, and $K_3$ can be determined by the minimization of the total mean-field energy. Our energy minimization gives $K_1=\pm K_3$ and $K_2=0$, which leads to the fully-gapped $d_{x^2-y^2}\pm i d_{xy}$ (abbreviated as $d+id$) pairing state. This mixture of the two $d$-wave pairings satisfies the requirement that the gap nodes should avoid the FS to lower the energy. Similarly, one can verify that the degenerate $p_x$ and $p_y$ pairing states will also mix into the fully-gapped $p_x\pm ip_y$ (abbreviated as $p+ip$) pairing state to lower the energy below $T_c$.

\end{widetext}

\end{document}